\documentclass[a4paper,fleqn,usenatbib]{mnras}

\usepackage{mathptmx}
\usepackage[T1]{fontenc}
\usepackage{ae,aecompl}

\usepackage{graphicx}	
\usepackage{amsmath}	
\usepackage{amssymb}	

\newcommand{\diff}{\mathrm{d}}
\newcommand{\Ms}{{\ensuremath{\mathrm{M}_{\odot}}}}
\newcommand{\Rs}{{\ensuremath{\mathrm{R}_{\odot}}}}
\newcommand{\Ls}{{\ensuremath{\mathrm{L}_{\odot}}}}
\newcommand{\Teff}{{\ensuremath{T_{\rm eff}}}}
\newcommand{\Mpy}{\Ms\,{\rm yr}{\ensuremath{^{-1}}}}
\newcommand{\fitm}{{\rm fit}\ensuremath{_M}}
\newcommand{\hii}{H~{\sc ii}}
\newcommand{\dm}{\ensuremath{\dot M}}
\newcommand{\dmc}{\ensuremath{\dot M_{\rm crit}}}
\newcommand{\mfin}{\ensuremath{M_{\rm fin}}}

\newcommand{\gva}{{\sc geneva}}
\newcommand{\kep}{{\sc kepler}}
\newcommand{\Td}{{\ensuremath{T_{\rm D}}}}

\newcommand{\Edd}{{\ensuremath{\Gamma_{\rm Edd}}}}
\newcommand{\tria}{{\ensuremath{3\alpha}}}
\newcommand{\sion}{{\ensuremath{S_{\rm ion}}}}
\newcommand{\slw}{{\ensuremath{S_{\rm LW}}}}
\newcommand{\sac}{{\ensuremath{s_{\rm ac}}}}
\newcommand{\guig}{\textquoteleft}
\newcommand{\guid}{\textquoteright}

\title[Supermassive Pop III Stars
]{The Evolution of Supermassive Population III Stars
}

\author[L. Haemmerl\'{e} et al.]{Lionel Haemmerl\'{e}$^{1}$\thanks{E-mail: Lionel.Haemmerle@unige.ch}, 
T.~E. Woods$^{2}$, 
Ralf S. Klessen$^{1,4}$, 
Alexander Heger$^{2}$, 
\newauthor Daniel J. Whalen$^{3}$\\
$^{1}$Universit\"at Heidelberg, Zentrum f\"ur Astronomie, Institut f\"ur Theoretische Astrophysik, Albert-Ueberle-Str. 2,
D-69120 Heidelberg, Germany\\
$^{2}$Monash Centre for Astrophysics, School of Physics and Astronomy, Monash 
University, VIC 3800, Australia\\
$^{3}$Institute of Cosmology and Gravitation, University of Portsmouth, Dennis Sciama Building, Portsmouth PO1 3FX, UK\\
$^{4}$Interdisziplin\"ares Zentrum f\"ur wissenschaftliches Rechnen der Universit\"at Heidelberg,
Im Neuenheimer Feld 205, D-69120 Heidelberg, Germany\\
}

\date{Accepted XXX. Received YYY; in original form ZZZ}

\pubyear{2017}

\begin{document}
\label{firstpage}
\pagerange{\pageref{firstpage}--\pageref{lastpage}}
\maketitle

\begin{abstract}
Supermassive primordial stars forming in atomically-cooled halos at $z \sim15-20$
are currently thought to be the progenitors of the earliest quasars in the Universe.
In this picture, the star evolves under accretion rates of 0.1 -- 1~\Mpy\
until the general relativistic instability triggers its collapse to a black hole at masses of $\sim10^5$~\Ms.
However, the ability of the accretion flow to sustain such high rates depends crucially on the photospheric properties of the accreting star,
because its ionising radiation could reduce or even halt accretion.
Here we present new models of supermassive Population III protostars accreting at rates 0.001 -- 10~\Mpy,
computed with the \gva\ stellar evolution code including general relativistic corrections to the internal structure.
We use the polytropic stability criterion to estimate the mass at which the collapse occurs,
which has been shown to give a lower limit of the actual mass at collapse in recent hydrodynamic simulations.
We find that at accretion rates higher than 0.001~\Mpy\ the stars evolve as red, cool supergiants
with surface temperatures below $10^4$~K towards masses $>10^5$~\Ms,
and become blue and hot, with surface temperatures above $10^5$~K, only for rates $\lesssim0.001$~\Mpy.
Compared to previous studies, our results extend the range of masses and accretion rates
at which the ionising feedback remains weak, reinforcing the case for direct collapse as the origin of the first quasars.

\end{abstract}





\section{Introduction}
\label{sec-in}

The properties and evolution of supermassive stars (SMS), with masses $\gtrsim$ 10$^4$ \Ms, have been studied since the early 1960s
\cite[e.g.,][]{osaki1966,unno1971,appenzeller1972a,appenzeller1972b,fricke1973,shapiro1979,fuller1986}.
But the existence of such stars has only recently been suspected to be necessary to explain the formation of quasars by $z \gtrsim$ 7,
such as ULAS J1120+0641, a $2 \times 10^9$~\Ms\ black hole at $z =$ 7.1 \citep{mortlock2011}
and SDSS J010013.02+280225.8, a $1.2\times10^{10}$ \Ms\ black hole at $z=6.3$ (\citealt{wu2015}; see \citealt{smidt2017}).
The origin of these supermassive black holes (SMBHs) may not have been $10-500$~\Ms\ Population III (Pop III) star BHs at $z\sim20-25$
because they might not have achieved the rapid, sustained growth needed to exceed $10^9$~\Ms\ by $z\gtrsim7$
\citep{whalen2004,alvarez2009,park2011,whalen2012}.
Supercritical accretion by Pop III star BHs could allow them to grow to such masses at early times even with limited duty cycles
\citep{volonteri2015,inayoshi2016a,sakurai2016b,pezzulli2016},
but it is not known if such processes operate in primordial accretion discs or for the times required to produce massive seeds.
The seeds of the first quasars may instead have been $10^4-10^5$~\Ms\ BHs that formed via direct collapse.

In this picture, a primordial halo forms in close proximity to nearby star-forming regions with strong Lyman-Werner (11.18 - 13.6 eV) UV
and H$^-$ photodetachment (>~0.755~eV) fluxes that sterilise the halo by effectively destroying the main coolant, molecular hydrogen H$_2$
(\citealt{agarwal2012,dijkstra2014,agarwal2016b}; but see also \citealt{inayoshi2012,inayoshi2015b}).
Due to the absence of H$_2$ molecules, the gas temperature rises to $10^4$~K,
preventing fragmentation and star formation before the halo's mass reaches a few $10^7$~\Ms.
At such masses, the halo gas finally becomes gravitationally unstable and begins to contract towards the centre with very high accretion rates of 0.1 -- 10~\Mpy,
forming a $10^4-10^5$~\Ms\ star in less than the lifetime of the star on the main sequence \citep[e.g.][]{latif2013d,latif2013e,becerra2015}.
The dynamics of these flows on the smallest scales are not yet fully understood,
but in the simulations performed to date a massive line-cooled disc forms that rapidly feeds the growth of a single object at its centre.
Fragmentation, if it occurs, is minor and the clumps mostly spiral into the central object \citep{regan2014,inayoshi2014b,becerra2015}.
It is expected that stars in this mass range will collapse directly to BHs without exploding, with masses equal to the progenitors
due to the inefficiency of radiative mass losses in metal-free stars.
An observational candidate for a direct collapse black hole (DCBH) has now been found, CR7, a Ly-$\alpha$ emitter at $z=6.6$
\citep{sobral2015,pallottini2015,hartwig2016,agarwal2016b}.
Current models favor the BH interpretation of CR7 over a Pop III starburst \citep{pallottini2015,hartwig2016,agarwal2016b}.

A number of studies have recently examined the evolution of supermassive Pop III protostars growing by accretion at the high rates expected for atomically-cooled haloes.
\citet{hosokawa2013} followed the growth of such objects up to $\sim10^5$~\Ms\ for different constant accretion rates
and found that for rates $\gtrsim0.1$~\Mpy\ the protostars remain red and cool until they reach a few 10$^4$ \Ms.
\citet{sakurai2015} studied the evolution of supermassive Pop III stars in clumpy accretion scenarios
and suggest that the protostar could intermittently become blue and hot at low masses but eventually evolved onto a redder, cooler track.
But the codes used in these studies did not include the general relativistic (GR) corrections to hydrostatic equilibrium,
and thus the runs were stopped before the stellar mass exceeds $10^5$~\Ms.
Indeed, above this mass, the GR effects are expected to become significant,
in particular by triggering the collapse into a black hole through the GR instability \citep{iben1963,chandrasekhar1964}.
\cite{umeda2016} included the post-Newtonian corrections to their models,
computing the internal structure of stars accreting at rates 0.1~--~10~\Mpy\ towards masses of $1-8\times10^5$~\Ms.
At these masses, the criterion of \cite{chandrasekhar1964} for GR stability, based on the assumption of polytropic structures, indicates instability.
However, no evolutionary tracks were displayed in this study, and the properties of the radiative feedback in their models,
in particular in the highest mass-range, are not available.

In a first paper \citep{woods2017}, we modeled the growth of supermassive Pop III stars at accretion rates of 0.01 - 10 \Mpy\ with the \kep\ stellar evolution code.
The \kep\ code includes a self-consistent treatment of the hydrodynamics, taking into account the post-Newtonian correction,
so that the collapse can be followed without the use of any external criterion.
We found that the mass at collapse varied from $\sim7.5\times10^4$ to $3.2\times10^5$ \Ms.
However, numerical difficulties in the integration of the atmosphere in case of accretion did not allow us to study the photospheric properties of our models,
and to establish the evolutionary tracks and ionising effect of the radiation field.

The actual growth rates of supermassive Pop~III stars may be crucially dependent on their internal structure and surface temperature.
If the star is red and cool, accretion proceeds at rates set by the cosmological flows assumed in previous evolution models.
But if the star becomes compact, blue and hot its ionising radiation could reduce accretion and the final mass of the star.
The accretion geometry is also critical: the ram pressure of spherical inflows at the rates previously studied
would almost certainly prevent radiation from the star from ionising the flows even if it is blue and hot \citep{johnson2012}.
Accretion through a disc, which is more likely, could result in bipolar radiation breakout that disperses the flows
\citep[e.g.,][]{hosokawa2011b,hirano2014}.
Clumpy accretion can likewise result in hot protostars that suppress their own growth at early times \citep{sakurai2015}.

In the present work, we re-examine the evolution of supermassive primordial protostars accreting at high rates (0.001 -- 10~\Mpy).
We present models computed with the \gva\ stellar evolution code, and describe the properties of their internal structures and evolutionary tracks.
We focus on the ionising properties of the radiation field, in order to evaluate the potential of these stars to regulate their own growth.
In addition, we use the polytropic criterion to study the development of the GR instability in the stellar interior,
expected to trigger the collapse of the protostar into a black hole.
In Sect.~\ref{sec-ge} we describe our \gva\ models.
In Sect.~\ref{sec-mod} we examine their interior structure, and surface properties.
In Sect.~\ref{sec-dis} we study how these surface properties depend on the treatment of their external layers,
we estimate the mass at which the GR instability triggers the collapse into a black hole, and we compare our results with those of previous studies.
We summarise our conclusions in Sect.~\ref{sec-out}.

\section{Numerical method}
\label{sec-ge}

The \gva\ code is a one-dimensional hydrostatic stellar evolution code
that numerically solves the four usual equations of stellar structure (e.g.~\citealt{eggenberger2008}) with the Henyey method.
The energy generation rate includes both nuclear reactions and gravitational contraction,
opacities are derived from the OPAL tables \citep{iglesias1996},
and convection is approximated by mixing-length theory.
A general description of the code for the case without accretion is given in \cite{eggenberger2008}.

Accretion has recently been implemented in the \gva\ code as described in \cite{haemmerle2013,haemmerle2016a},
and we recall here only the main ideas.
The accretion rate is a free parameter, fixed externally.
Here we consider the following constant rates:
\begin{equation}
\dm=0.001,\,0.01,\,0.1,\,1,\,10\ \Mpy
\label{eq-dm}\end{equation}
Since the code is hydrostatic, we model only the stellar interior, without the accretion shock.
Moreover, the code does not include any contribution to the luminosity from the accretion energy,
and we assume that the entropy of the accreted material is the same as that of the stellar surface.
This assumption corresponds to accretion onto the star through a disc,
for which any entropy excess can be efficiently radiated away in the polar direction before being advected in the stellar interior
(cold disc accretion, \citealt{palla1992,hosokawa2010}).

GR effects are expected to be important in SMS, and to account for them
we apply the first order post-Newtonian Tolman-Oppenheimer-Volkoff (TOV) correction in the equation of hydrostatic equilibrium.
We replace the Newtonian gravitational constant $G$ by
\begin{equation}
G_{\rm rel}=G\left(1+{P\over\rho c^2}+{2GM_r\over rc^2}+{4\pi Pr^3\over M_rc^2}\right)
\label{eq-g}\end{equation}
where $P$ is the pressure, $\rho$ the mass density, $c$ the speed of light and $M_r$ the mass enclosed in a radius $r$.
This approximation to GR is the same as that in \kep\ (\citealt{fuller1986}).

We emphasise that, as usual, the outer regions of the star are not included in the calculation of the stellar interior.
For numerical stability, one has to neglect the production and absorption of energy in the layers with $M_r/M>\fitm$, where \fitm\ is fixed externally.
In these layers, the structure equations are integrated assuming a flat luminosity profile.
Decreasing \fitm\ favours numerical convergence, while increasing it improves physical consistency.
In all the models described in the present work, we fix a value of \fitm\ which is constant during the evolution.
Depending on the models, we consider either $\fitm=0.999$ or $\fitm=0.99$.
The consequences of this assumption are discussed in Sect.~\ref{sec-dis-fm}.

Convective zones are determined according to the Schwarzschild criterion.
For numerical stability, we do not include any overshooting.
The consequences of these choices are discussed in Sect.~\ref{sec-dis-lit}.

\section{Models}
\label{sec-mod}

\subsection{Initial setups}
\label{sec-mod-ini}

Accretion at high rates onto low-mass hydrostatic cores makes numerical convergence difficult.
Thus for 0.1, 1 and 10~\Mpy\ we initialise our models with a mass of $M_{\rm ini}=10$~\Ms,
while for 0.01 and 0.001 \Mpy\ we take $M_{\rm ini}=2$~\Ms.
The chemical composition of the initial models is homogeneous, with a hydrogen mass fraction of $X=0.7516$,
a helium mass fraction of $Y=0.2484$, and a metallicity $Z=1-X-Y=0$.
We include deuterium with a mass fraction of $X_2=5\times10^{-5}$ (\citealt{bernasconi1996a,behrend2001,haemmerle2016a}).
The chemical composition of the accreted material is identical to that of the initial protostellar seeds.
The initial structures correspond to polytropes with $n\simeq3/2$, with flat entropy profiles, so that the stars start their evolution as fully convective objects.
The central temperatures are $4.1\times10^5$~K and $6.6\times10^5$~K for the 10~\Ms\ and 2~\Ms\ initial models, respectively,
which is below the temperature required for D-burning ($\Td\simeq1-2\times10^6$~K).
We choose the initial time-step in order to ensure that the mass accreted in the first time-step does not exceed 0.1 \Ms.
As a consequence, the initial time-step depends on the accretion rate, and is given by $\diff t=$ 0.1 \Ms/\dm.
We take $\fitm=0.999$ as a fiducial value, except for 0.001 \Mpy.
The motivations and consequences of this choice are discussed in Sect.~\ref{sec-dis-fm}.
For reasons of numerical stability, we do not include the GR correction in the model at $\dm=0.001$ \Mpy.
This model never exceeds $10^4$ \Ms\ significantly, so that we expect GR effects to be negligible in this case.

\subsection{Evolutionary tracks and internal structures}
\label{sec-mod-evol}

\begin{figure}\begin{center}\includegraphics[width=0.45\textwidth]{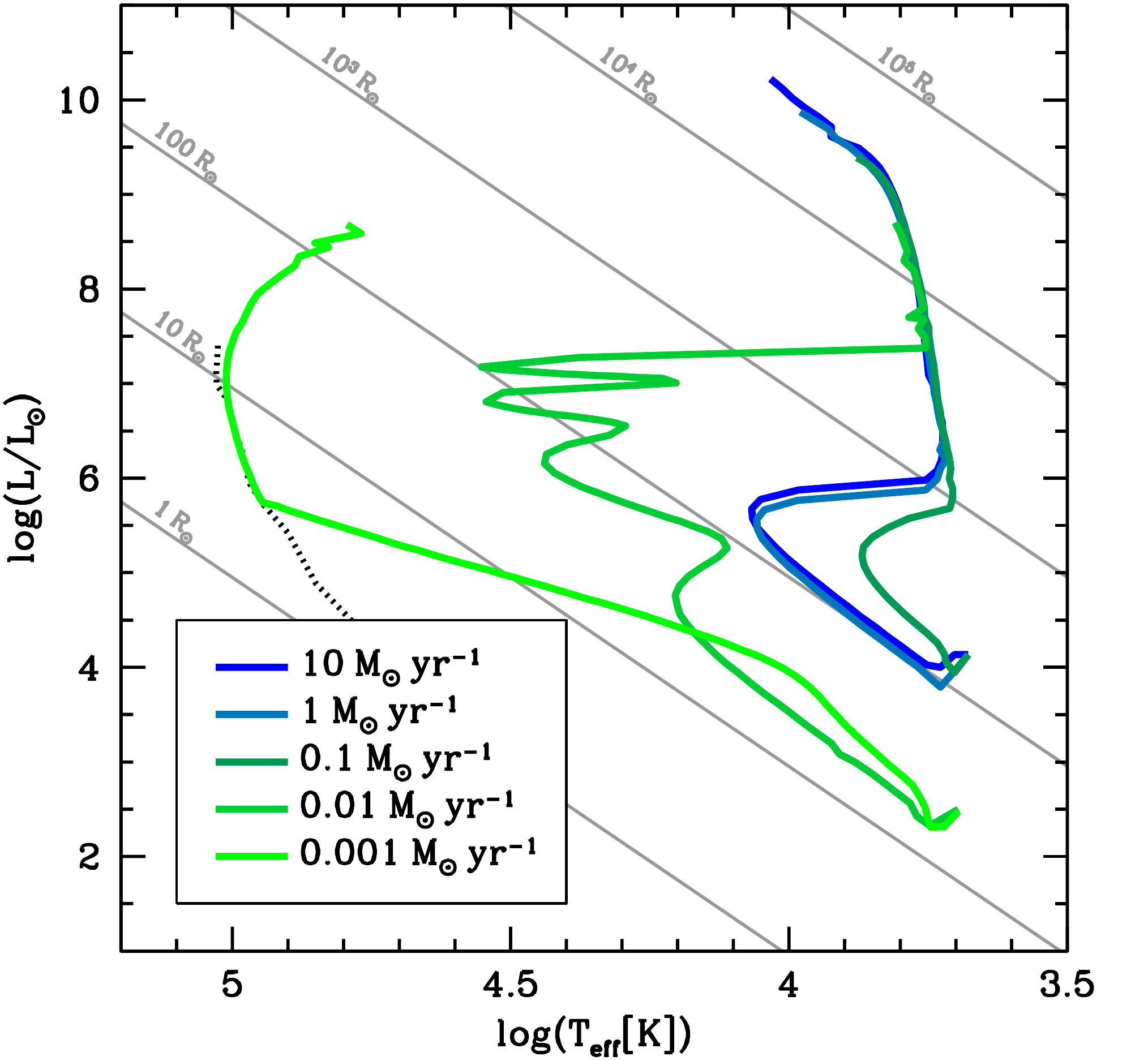}\end{center}
\caption{Evolutionary tracks on the HR diagram for the models at the indicated accretion rates.
The grey straight lines indicate the stellar radius, and the black dotted curve is the ZAMS of \citealt{schaerer2002}.}
\label{fig-hr}\end{figure}

The evolutionary tracks on the Hertzsprung-Russel (HR) diagram are shown on Fig.~\ref{fig-hr} for the five accretion rates.
For all the models, the luminosity increases monotonically as the stellar mass grows by accretion, except in the very early evolution.
The mass-luminosity relation is nearly independent of the accretion history (Fig.~\ref{fig-tl}, lower panel).
But the evolution of the effective temperature differs significantly between models at various rates.
After an adjustment phase, the tracks converge towards two distinct asymptotic regimes in the HR diagram:
the high-\dm\ regime ($\dm\gtrsim0.01$\,\Mpy) gives a nearly vertical track in the red, along the Hayashi limit,
while the low-\dm\ regime ($\dm<0.01$\,\Mpy) leads to the blue, along the Zero-Age Main Sequence (ZAMS).
In the high-\dm\ regime, the effective temperature is locked around 5000 -- 6000 K,
and thus never exceeds $10^4$ K before the luminosity reaches $10^{10}$~\Ls.
The star is bloated up, with a radius larger than 1000~\Rs, as a \guig red supergiant protostar\guid\ \citep{hosokawa2012a}.
In the low-\dm\ regime, the track evolves immediately towards the blue, approaching $\Teff\simeq10^5$~K before the luminosity exceeds $10^6$~\Ls.
The increase in effective temperature is only stopped when the star reaches the ZAMS and stops contracting.
Thus the location of the star on the HR diagram is confined between two limits: the Hayashi limit in the red and the ZAMS in the blue.
Each of these limits corresponds to the asymptotic track of our models according to their accretion regime, low- or high-\dm,
i.e. depending if the rate is above or below a critical value $\dmc\sim0.005$~\Mpy.
Notice that the Hayashi limit reflects the physics at the stellar surface, while the ZAMS limit reflects the physics at the centre.
Models at $\dm=0.001$ \Mpy\ or with $\dm\geq0.1$ \Mpy\ converge to their asymptotic track relatively early,
before the luminosity exceeds significantly $10^6$~\Ls.
For the intermediate case $\dm=0.01$~\Mpy, the track remains longer between the two asymptotic limits,
showing oscillations in \Teff\ in the range 10\,000 -- 30\,000 K.
Convergence towards the Hayashi limit occurs eventually when the luminosity has grown to $2\times10^7$~\Ls.

\begin{figure}\begin{center}\includegraphics[width=0.49\textwidth]{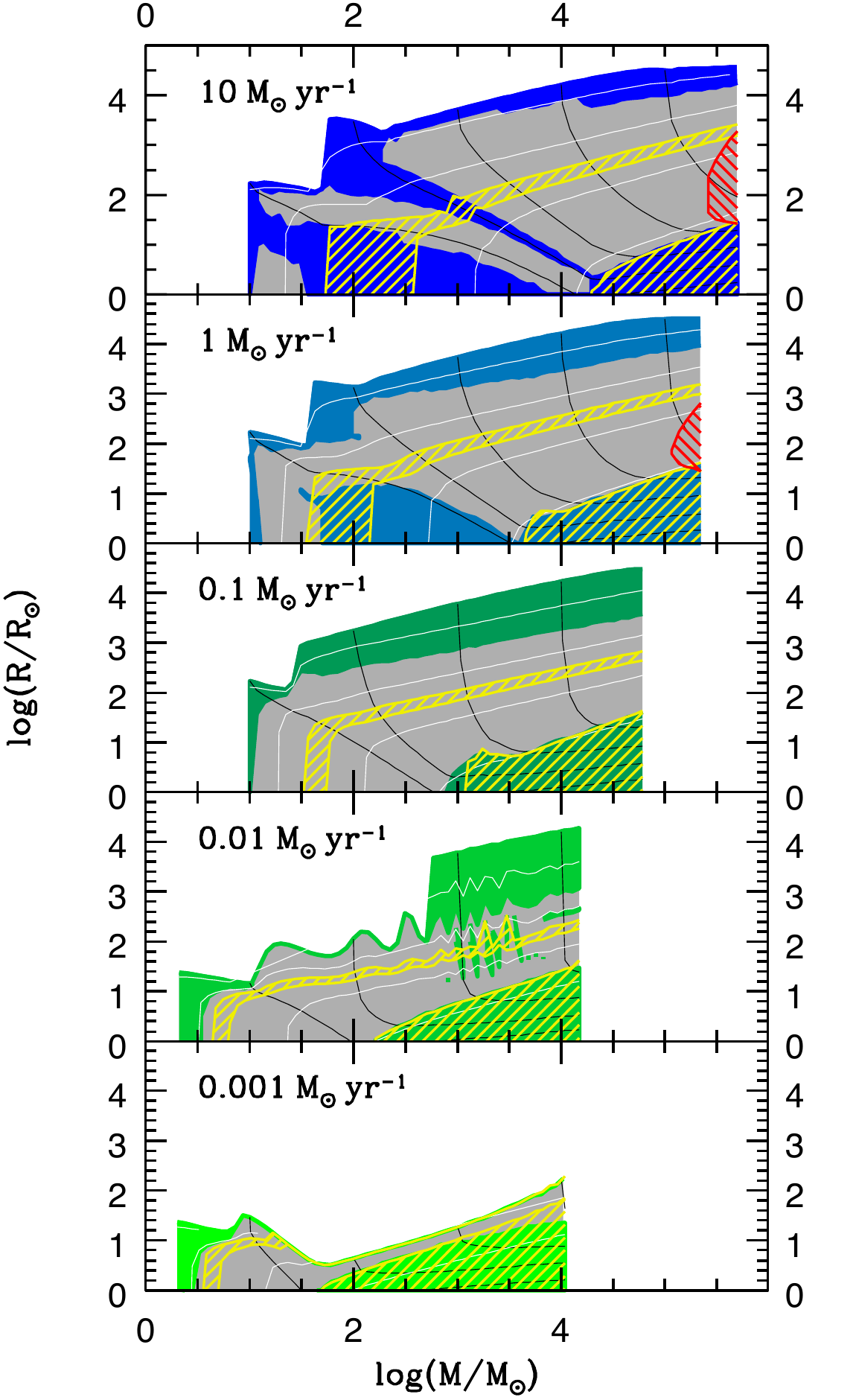}\end{center}
\caption{Internal structures of the models for the indicated accretion rates.
On each panel, the upper curve is the stellar radius, the blue and green areas indicate convective zones,
and the grey areas indicate radiative transport.
The yellow hatched areas correspond to D- and H-burning,
and the red hatched areas indicate the GR instability according to the polytropic criterion of Eq.~(\ref{eq-crit}) with $n=3$.
The black curves indicate the Lagrangian layers of $\log(M_r/\Ms)=$ 1, 2, 3, 4 and 5,
and the white ones are isotherms of $\log(T\rm[K])=$ 5, 6, 7 and 8.}
\label{fig-st}\end{figure}

\begin{figure}\begin{center}\includegraphics[width=0.45\textwidth]{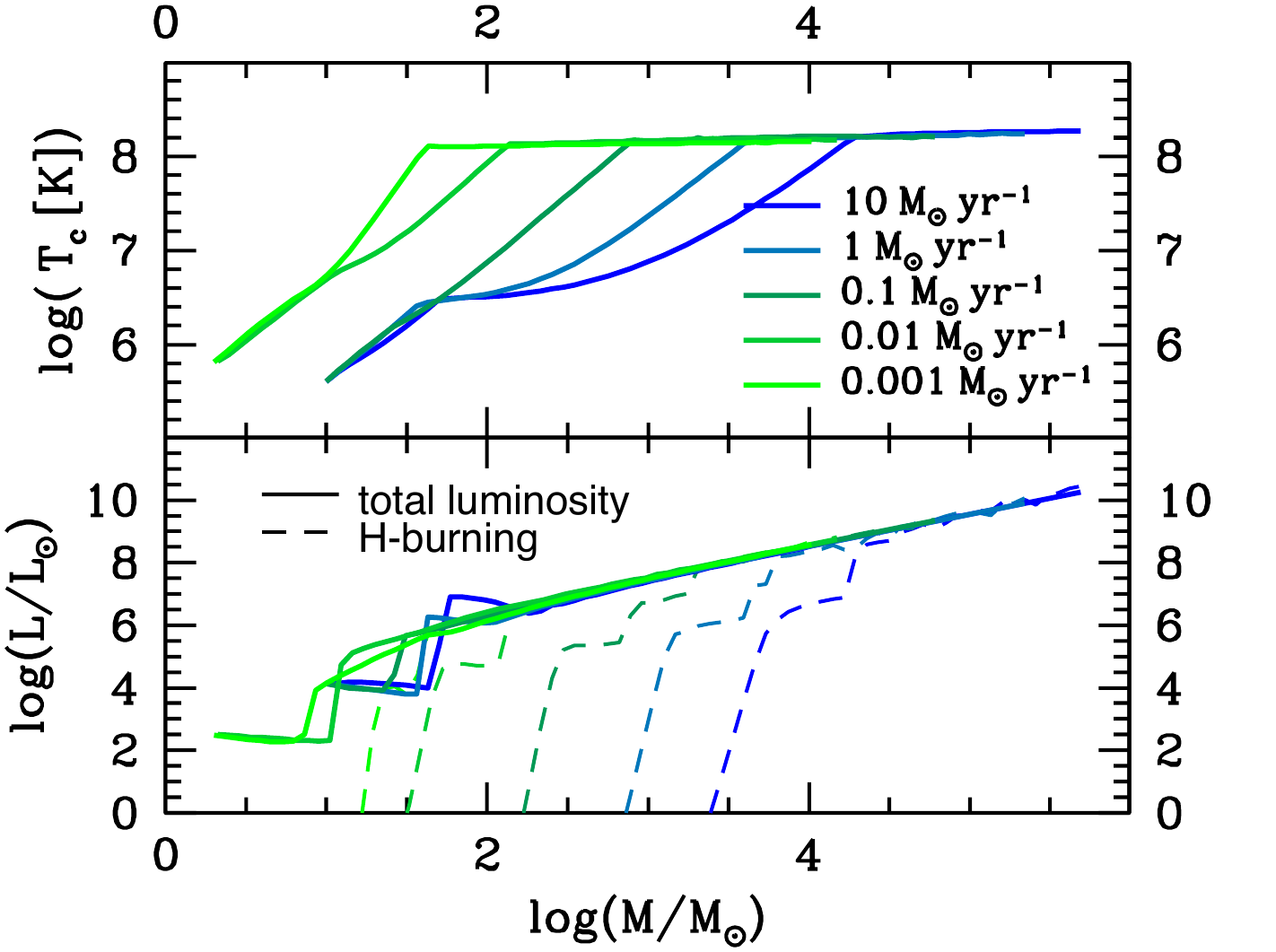}\end{center}
\caption{Evolution of the central temperature (upper panel) and of the luminosity (lower panel)
as a function of the stellar mass, for models with the indicated accretion rates.
On the lower panel, the solid lines indicate the total luminosity, and the dashed lines the contribution from H-burning only.}
\label{fig-tl}\end{figure}

\begin{figure}\begin{center}\includegraphics[width=0.4\textwidth]{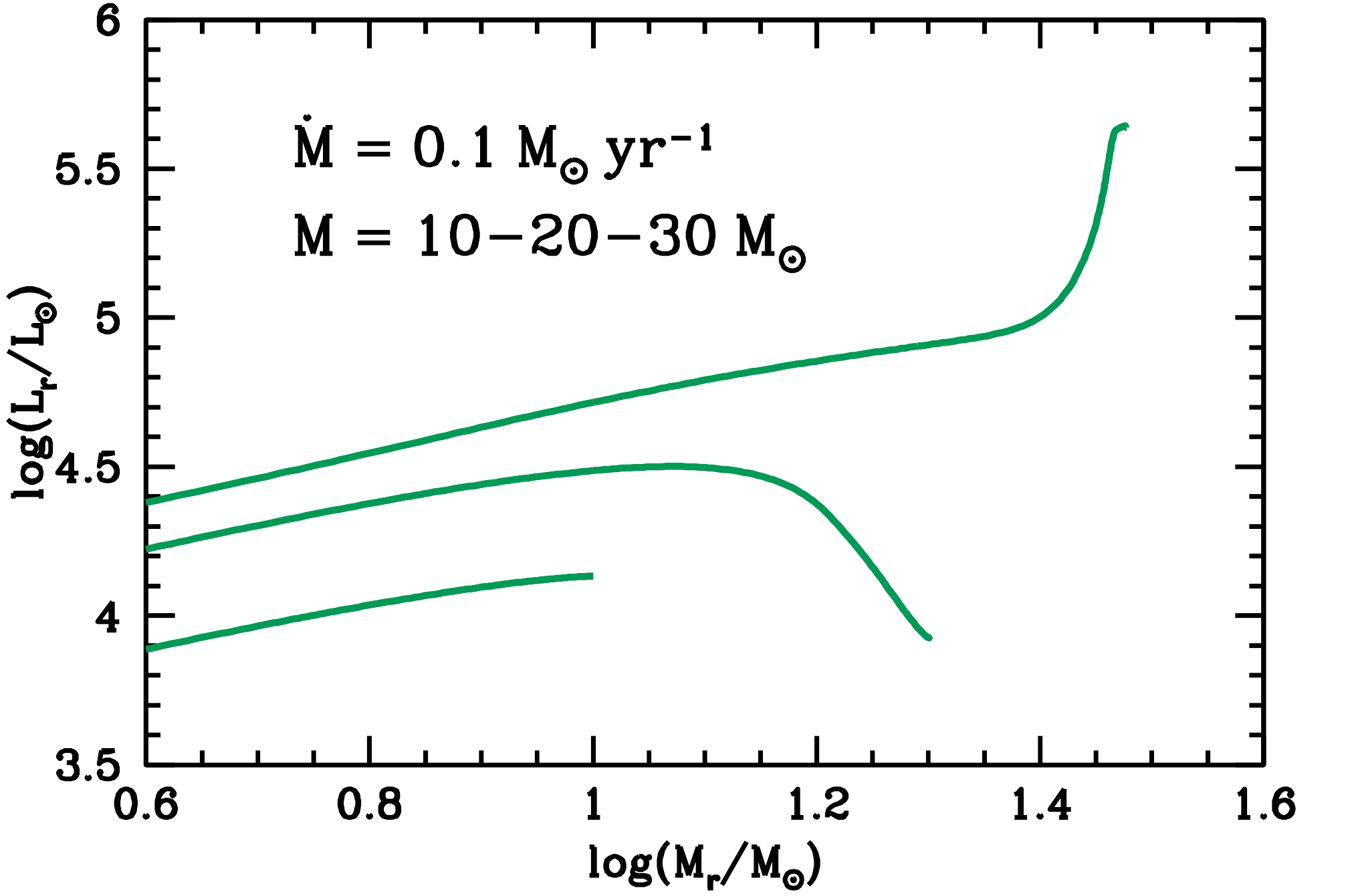}\end{center}
\caption{Luminosity wave at 0.1 \Mpy.
The three curves show the internal luminosity profiles at $M=10$, 20 and 30~\Ms.
The first profile corresponds to the fully convective initial model.
At the second one, the radiative core has grown to 80\% of the total stellar mass, but the stellar radius is still decreasing ($R=118$~\Rs).
In the last profile, the luminosity wave has reached the surface, $\diff L_r/\diff M_r>0$ everywhere, and the radius has increased to 821~\Rs.}
\label{fig-lwav}\end{figure}

The internal structure of these models is illustrated in Fig.~\ref{fig-st}.
In addition, Fig.~\ref{fig-tl} shows the evolution of their central temperatures and surface luminosities.
All five models start with a fully convective structure and a central temperature $T_c<\Td$.
The star takes its energy from Kelvin-Helmholtz (KH) contraction,
loosing entropy ($\diff L_r/\diff M_r=-T\diff s/\diff t>0$) and increasing $T_c$.
As $T_c$ increases, the opacity in the centre becomes low enough for a radiative core to form and grow in mass.
The growth of the radiative core follows the isotherms,
which reflects the fact that the transition from convection to radiation is an effect of the temperature increase, through the opacity.
Only for $\dot M=10$~\Mpy, intermediate convective zones survive in between the radiative regions.
Since these zones correspond to the Lagrangian layers of the initial seed, we expect their properties to reflect the choice of the initial structure.
The reduction of opacity in the central regions produces an increase in the internal luminosity in radiative regions ($\diff L_r/\diff M_r>0$).
The entropy produced in these regions is absorbed by the cold external convective layers with high opacity ($\diff L_r/\diff M_r<0$).
As the internal temperature increases, the boundary between these two regions moves outwards in mass.
When this \guig luminosity wave\guid\ \citep{larson1972,hosokawa2010} reaches the stellar surface,
the radius increases abruptly, by more than one order of magnitude for $\dm\ge1$~\Mpy, by a factor of a few for $\dm\leq0.01$~\Mpy.
Fig.~\ref{fig-lwav} shows the evolution of the luminosity wave for the model at 0.1~\Mpy.

In the while, $T_c>\Td\simeq1-2\times10^6$ K (Fig.~\ref{fig-tl}, upper panel; see also the isotherm of $10^6$ K on Fig.~\ref{fig-st}),
and deuterium starts burning:
in the radiative core for $\dm\leq0.1$~\Mpy, in the central convective zones for $\dm\geq1$~\Mpy(Fig.~\ref{fig-st}).
Once deuterium is exhausted in the centre, the D-burning region moves outwards in mass (shell D-burning),
following the isotherms (Fig.~\ref{fig-st}).
The convective core that formed in the high-\dm\ models survives the D exhaustion,
but remains confined in the same Lagrangian layers that correspond to the initial seed, and thus contracts with it.
Notice that neither this convective zone nor the plateau in central temperature visible on Fig.~\ref{fig-tl} (at $30\,\Ms<M<100\,\Ms$)
are due to D-burning, as one could naively believe.
Test computations without deuterium confirmed that this features appear in any case.
Actually, our computations show that D-burning has no significant impact on the stellar structure of the models described here.

After the luminosity wave has reached the surface, all the layers of the star loose entropy ($\diff L_r/\diff M_r=-T\diff s/\diff t>0$).
For low enough accretion rates ($\dm\sim0.001$~\Mpy), the stellar radius decreases as a consequence,
despite the new mass which is continuously accreted, and the star can contract towards the ZAMS.
For high \dm\ however, the entropy losses are not efficient enough to make the stellar radius to decrease.
Despite the contraction of all the layers, the new material that lands on the stellar surface makes the stellar radius
to increase monotonically for the rest of the evolution.

\begin{figure}\begin{center}\includegraphics[width=0.4\textwidth]{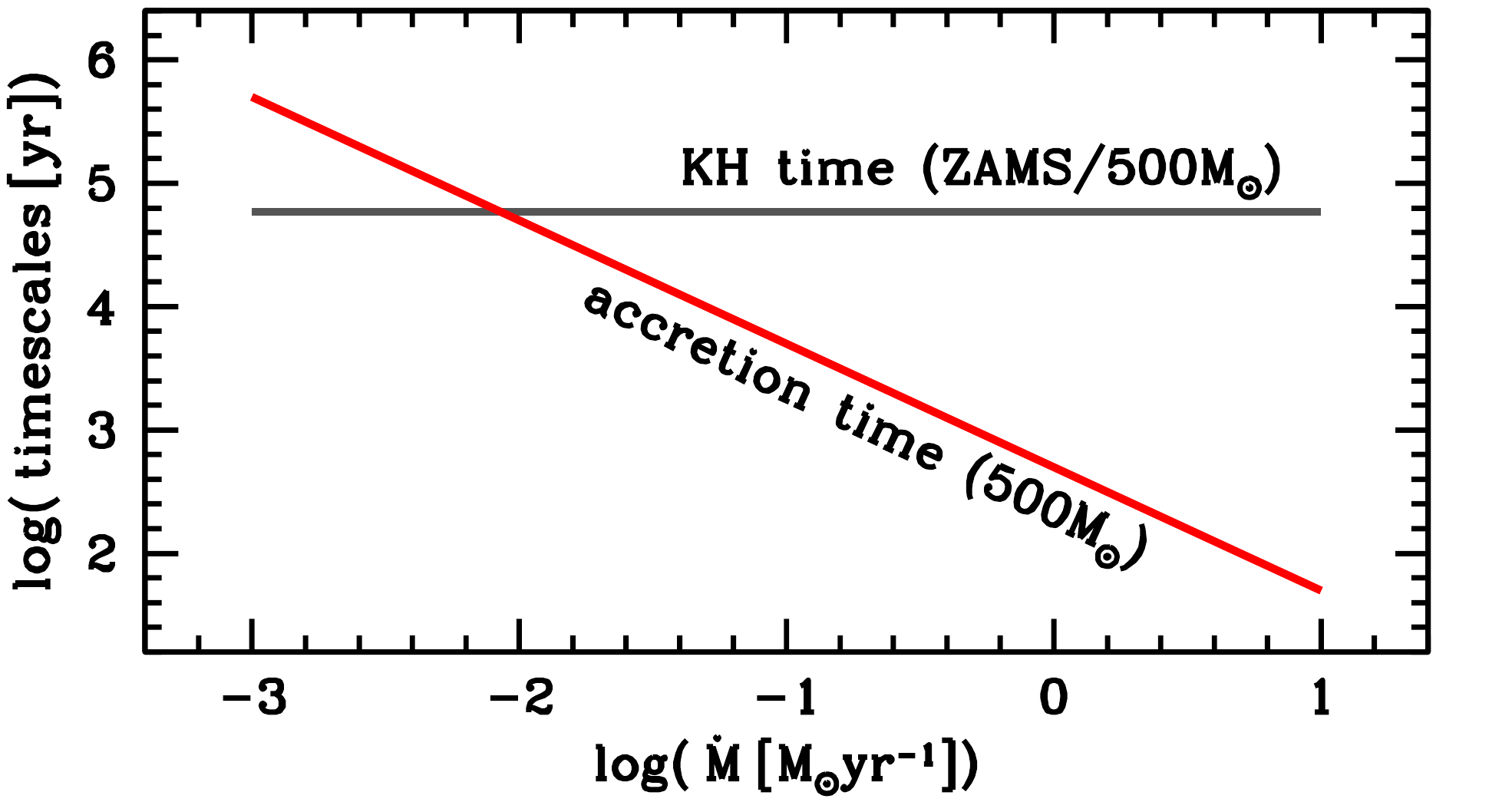}\end{center}
\caption{Timescales balance: comparison between the timescales for accretion and KH contraction at $M=500$\,\Ms.
The KH time is computed using the ZAMS radius and luminosity \citep{schaerer2002}.}
\label{fig-time}\end{figure}

In order to illustrate the origins of this difference between the low- and high-\dm\ regime,
we plot on Fig.~\ref{fig-time} the timescales for accretion and KH contraction.
The KH time gives the timescale for thermal adjustment in the stellar interior,
and indicates the time it takes for the star to contract towards the ZAMS if accretion stops.
We use $t_{\rm KH}=GM^2/RL$ for $M=500$\,\Ms, with $R$ and $L$ from the ZAMS of \cite{schaerer2002}.
The accretion time is simply $M/\dot M$ for the same value of the mass, and gives the timescale of the evolution in the accretion phase.
The timescales balance changes according to the accretion rate.
For $\dm=0.001$~\Mpy, the KH time is shorter by one order of magnitude than the accretion time,
while for $\dm\ge0.1$~\Mpy\ it is longer by one order of magnitude or more (three orders of magnitude for 10~\Mpy).
For the intermediate case $\dm=0.01$~\Mpy, both timescales are similar.
As a consequence, in the low-\dm\ regime, the star has the time to contract towards the ZAMS before its mass increases significantly,
while in the high-\dm\ regime the mass increases too fast, and the stellar radius grows.

After this point, the behaviour of the various models remains qualitatively different depending on the regime, low- or high-\dm.
In the low-\dm\ regime, the internal structure is qualitatively similar to that of stars at present days \citep[see e.g.~][]{haemmerle2016a}.
However, due to the lower opacity, the star remains more compact and convective zones are thiner.
As a consequence, shell D-burning occurs in the radiative core instead of the convective envelope.
In the model at 0.001~\Mpy, the swelling leads to a maximum radius of 32 \Rs\
(instead of 48.6 \Rs\ in the present-day case, see model CV2 in \citealt{haemmerle2016a}).
Then, at $M=8.5$~\Ms, the star becomes fully radiative and contracts.
The central temperature increases (Fig.~\ref{fig-tl}, upper panel), triggering the \tria-reaction.
The energy produced by the \tria-reaction remains negligible.
The total luminosity is dominated by the gravitational contribution at this stage (KH contraction, $\diff L_r/\diff M_r\simeq-T\diff s/\diff t$).
However, the reaction produces enough $\rm^{12}C$ in the centre in order to trigger the CNO cycle,
which becomes the dominant energy source for the rest of the evolution (Fig.~\ref{fig-tl}, lower panel).
As a consequence, a convective core forms at $M=40$~\Ms\ and grows in mass (Fig.~\ref{fig-st}, lower panel),
while the central temperature is locked at $T_c=1.26\times10^8$ K due to the thermostatic effect of H-burning (Fig.~\ref{fig-tl}, upper panel).
The radius reaches a minimum of 3.3 \Rs\ at $M=50$~\Ms, and then grows continuously until the end of the computation,
according to the homologous relation $T_c\propto M/R$ and the fact that $T_c\simeq$~cst.
At $M=11\,650$ \Ms, numerical convergence becomes too difficult and we stop the computations, while H-burning is still proceeding.

In the high-\dm\ regime, the structure evolves in a qualitatively different way.
Due to the fast mass load at the surface, the radius can not contract and the star remains large.
This leads to low temperatures in the outer regions, which keep thus a high opacity and stay convective.
Below this convective envelope, intermediate convective zones appear.
A high accretion rate favours the formation and the development of these convective regions, because of the timescales balance:
the higher the rate, the shorter the time for the star to radiate the entropy contained in the deepest regions before reaching a given mass.
For $\dm=10$~\Mpy, in addition to the convective core described above, an intermediate convective region forms in the Lagrangian layers
that were accreted during the swelling (see the iso-mass of $100$~\Ms\ on the upper panel of Fig.~\ref{fig-st}).
This convective zone results from the accretion of entropy when the luminosity wave crosses the surface.
When the peak of the wave approaches the surface ($\diff L_r/\diff M_r=-T\diff s/\diff t<0$), the surface entropy increases suddenly.
Through the assumption of cold disc accretion, the specific entropy \sac\ of the material that is accreted increases then.
After the passage of the peak, the surface can radiate its entropy efficiently ($\diff L_r/\diff M_r=-T\diff s/\diff t>0$), and \sac\ decreases suddenly.
For the layers that are accreted at this stage, the decrease in the accreted entropy results in a negative entropy gradient
($\diff s/\diff M_r<0$), which drives convection.
Then, entropy is redistributed on a thermal timescale in the interior,
but this mechanism remains inefficient at high \dm\ because of the timescales balance between the KH and accretion times.
This is why this intermediate convective zone appears only in the model at $\dm>1$~\Mpy.
Despite the growth of the stellar radius, each Lagrangian layer contracts at this stage.
As a consequence, the central temperature increases until H-burning starts (Fig.~\ref{fig-tl}).
The physics of H-burning is the same as in the low-\dm\ regime:
the \tria-reaction is triggered first, and produces the $\rm^{12}C$ that allows the CNO-cycle to operate
as the dominant energy source for the rest of the evolution (Fig.~\ref{fig-tl}, lower panel).
A convective core forms (Fig.~\ref{fig-st}), and the central temperature is locked at $T_c\simeq1.25-2\times10^8$~K
by the thermostatic effect of H-burning (Fig.~\ref{fig-tl}, upper panel).
Notice that the higher the accretion rate, the higher the mass at which H-burning starts.
This is due to the timescale balance: the higher the rate, the shorter is the accretion time compared to the KH time
and thus the higher the mass accreted during the KH contraction towards the ZAMS.
Once the convective core has formed, the evolution proceeds in a regular way.
The stellar structure is made of three zones: the convective core, that grows in mass and radius along the isotherms ($T\simeq10^8$~K),
the convective envelope, that covers less than 10\% of the stellar mass (less than 2\% during most of the evolution),
and an intermediate radiative region in between.
While H burns in the convective core, D burns in a thin shell of the intermediate radiative zone, following the isotherms $T\simeq\Td$.
As the stellar mass grows by accretion, the stellar radius continues to increase monotonically.

Our models run until they reach different final masses, for reasons that are discussed below (Sect.~\ref{sec-dis-mfin}).
However, we notice that none of our models in the high-\dm\ regime shows a decrease of the stellar radius when the stellar mass exceeds $10^4$ \Ms.
For $\dm\geq1$~\Mpy, our models reach final masses of several $10^5$~\Ms, with a radius that is still growing.
As a consequence, the effective temperature remains lower than $10^4$ K until the stellar mass exceeds $3.5\times10^5$~\Ms.
This is in contrast with the results of previous studies, as discussed below (Sect.~\ref{sec-dis-lit}).

\begin{figure}\begin{center}\includegraphics[width=0.49\textwidth]{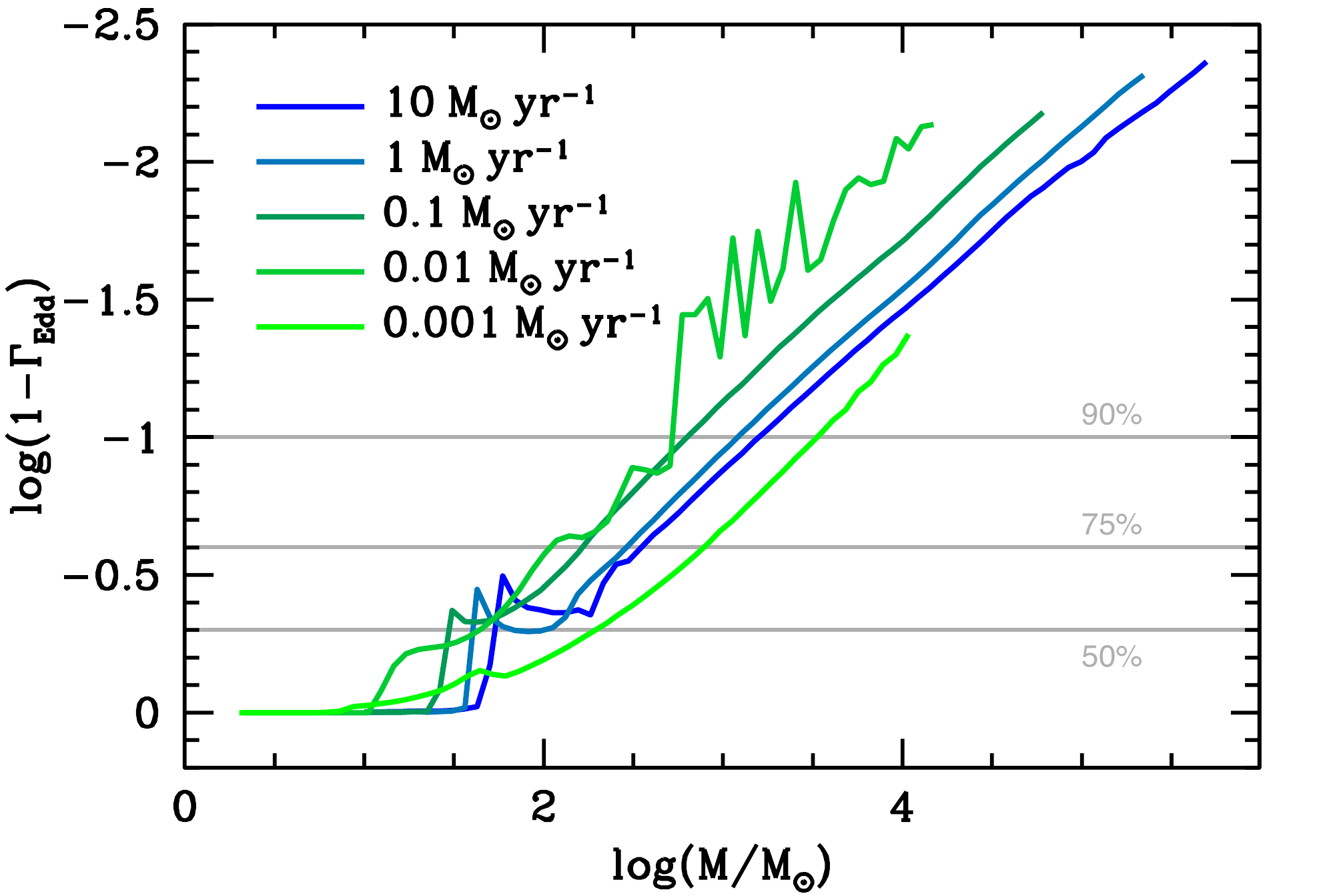}\end{center}
\caption{Evolution of the Eddington factor $\Edd=\nabla P_{\rm rad}/\nabla P$ at the surface
for the models with indicated accretion rates.
The values $\Edd=50$, 75 and 90\%\ are indicated by horizontal grey lines.}
\label{fig-edd}\end{figure}

In the intermediate case 0.01 \Mpy, after the swelling, the radiative star starts to contract towards the ZAMS as in the low-\dm\ regime.
But before it contracts significantly, in contrast to the 0.001 \Mpy\ case, the Eddington factor exceeds 50\%\ at the surface (Fig.~\ref{fig-edd}).
As a consequence, a second swelling occurs.
After several oscillations, at a mass of 600 \Ms, \Edd\ exceeds 90\%, the radius expands to $\simeq5000$~\Rs\
and the model converges towards the high-\dm\ regime.
Notice that in contrast to the models with $\dm\geq0.1$ \Mpy, the model at 0.01 \Mpy\ evolves to the red
because of the high radiation pressure (\Edd>90\%), and not because of the high gas pressure related to the high entropy content which cannot be radiated away.
In this intermediate case, oscillating between the two regimes, the contribution from radiation pressure is critical in determining the asymptotic behaviour.
Notice also that several intermediate convective zones survive between the convective core and the convective envelope in the model at 0.01 \Mpy.
They reflect the accretion history of entropy, in a similar way as the intermediate convective zone of the model at $\dm=10$~\Mpy.
However, at such low rate, thermal adjustment is efficient enough to redistribute the entropy before these zones join the convective core.

\subsection{Ionising feedback and Lyman-Werner flux}
\label{sec-mod-ion}

Our models show that the effective temperature of a star accreting at 0.001 -- 0.1 \Mpy\ depends sensitively on the rate.
To determine if the star can regulate accretion onto itself by radiative feedback,
we compute the number of ionising photons per second by integrating the black-body spectrum above the ionising energy:
\begin{eqnarray}
\sion&=&4\pi R^2\int\limits_{\scriptscriptstyle h\nu>13.6\rm\,eV}{F_\nu\over h\nu}\,d\nu \nonumber\\
&=&{8\pi^2R^2\over c^2h^2}\int\limits_{\scriptscriptstyle h\nu>13.6\rm\,eV}{(h\nu)^2\over e^{h\nu/k\Teff}-1}\,d\nu
\label{eq-ion}\end{eqnarray}
We compute also the Lyman-Werner (LW) flux (11.18 - 13.6 eV):
\begin{eqnarray}
\slw={8\pi^2R^2\over c^2h^2}\int\limits_{\scriptscriptstyle h\nu=11.18\rm\,eV}^{\scriptscriptstyle 13.6\rm\,eV}{(h\nu)^2\over e^{h\nu/k\Teff}-1}\,d\nu
\label{eq-lw}\end{eqnarray}

\begin{figure}\begin{center}\includegraphics[width=0.4\textwidth]{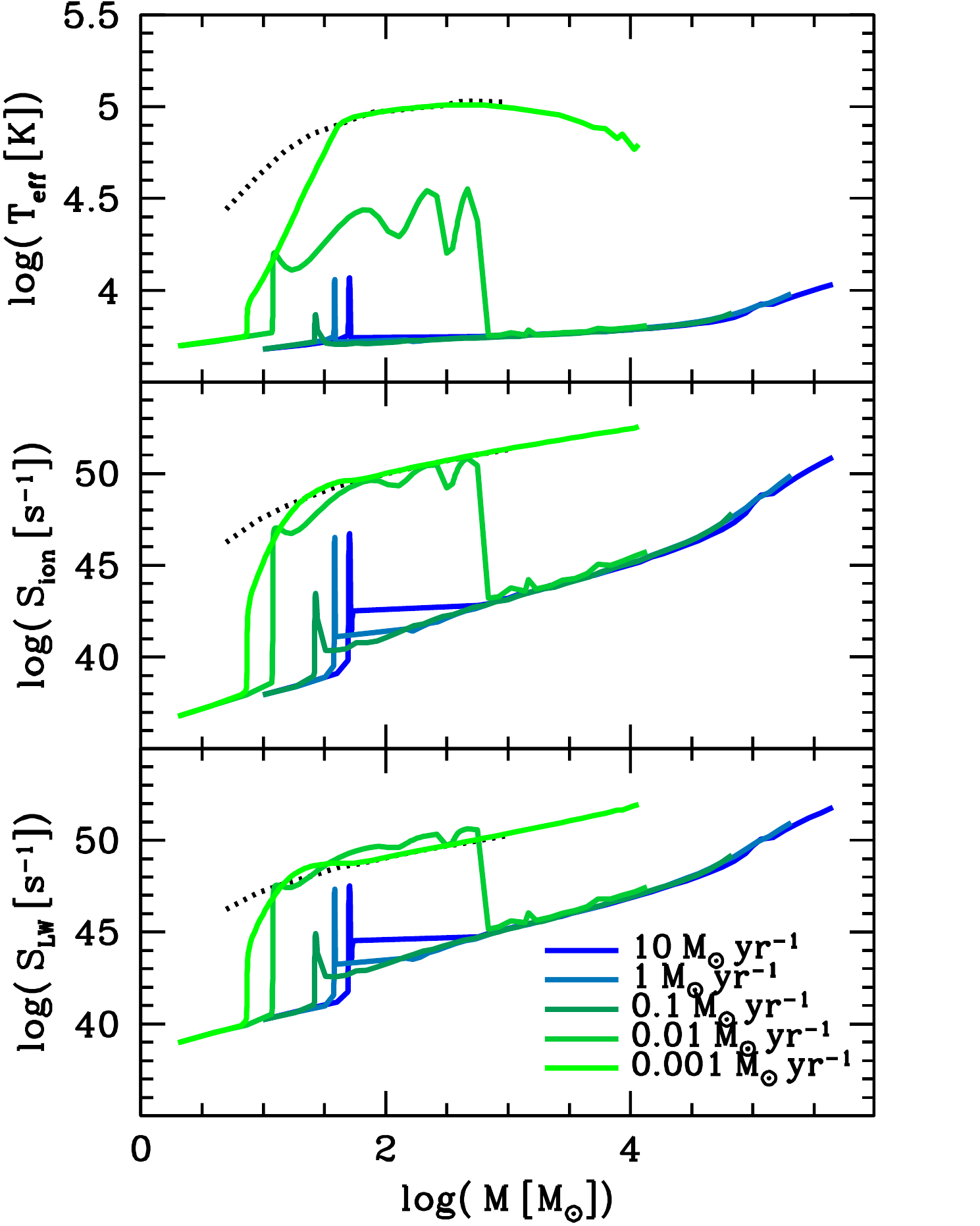}\end{center}
\caption{Evolution of the effective temperature (upper panel), ionising photon rate (middle panel)
and Lyman-Werner photon rate (bottom panel) as a function of the stellar mass
for the models with indicated accretion rates.
The black dotted lines correspond to the ZAMS of \citealt{schaerer2002}.}
\label{fig-ion}\end{figure}

The evolution of \sion\ and \slw\ as a function of the stellar mass is shown on Fig.~\ref{fig-ion} for the five accretion rates.
The two asymptotic limits in the evolutionary tracks reflect in two asymptotic limits in the ionising and LW fluxes.
In the low-\dm\ case, the ionising photon rate exceeds $10^{45}$\,s$^{-1}$ before the stellar mass reaches 20~\Ms.
This increase slows down only when \Teff\ reaches the ZAMS limit.
Then \sion\ grows slower, exceeding $10^{50}$\,s$^{-1}$ when the mass is $M>100$~\Ms.
In the high-\dm\ regime in contrast, after a short jump corresponding to the adjustment phase to the asymptotic behaviour,
the ionising photon rate remains lower than $10^{45}$\,s$^{-1}$ until $M=10^4$~\Ms,
and lower than $10^{50}$\,s$^{-1}$ until $M\simeq3\times10^5$~\Ms.
The fluxes grow slowly as the mass increases, following the Hayashi limit.
For the intermediate rate 0.01~\Mpy, the ionising and LW photon rates follow first the ZAMS limit.
The oscillations in effective temperature hardly impact the fluxes, because \Teff\ remains in the range 10\,000 -- 30\,000~K.
But when the model converges to the Hayashi limit, at $M=600$~\Ms, the fluxes switch to the high-\dm\ regime for the rest of the evolution.

\section{Discussion}
\label{sec-dis}

\subsection{Effect of changing the value of \fitm}
\label{sec-dis-fm}

As mentioned in Sect.~\ref{sec-ge}, the \gva\ code assumes that the energy generation rate is zero and the luminosity profile is flat
in the external layers of the star, those above a given value of $\fitm=M_r/M$.
In our models above, we used $\fitm=0.999$, except for the case $\dm=0.001$ \Mpy\ where we used $\fitm=0.99$.
Here we study the effect of changing the value of \fitm.
To that aim, we present models at $\dm>0.001$ \Mpy\ with $\fitm=0.99$, and a model at 0.001 \Mpy\ with $\fitm=0.999$.
Decreasing \fitm\ reduces physical consistency, but makes numerical convergence easier.
For $\dm<1$~\Mpy, models with $\fitm=0.99$ are started from the 2~\Ms\ initial protostellar seed, instead of the 10~\Ms\ one (Sect.~\ref{sec-mod-ini}).
For $\dm=0.001$~\Mpy, the model with $\fitm=0.999$ is started from the 10 \Ms\ seed.

\begin{figure}\begin{center}\includegraphics[width=0.49\textwidth]{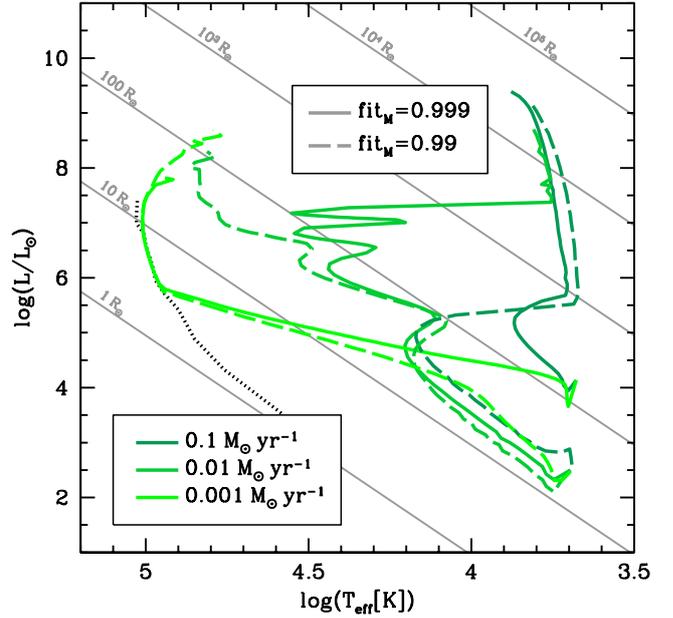}\end{center}
\caption{Evolutionary tracks of the models for the indicated accretion rates,
for $\fitm=0.999$ (solid lines) and 0.99 (dashed lines).
The grey straight lines indicate the stellar radius, and the black dotted curve is the ZAMS of \citealt{schaerer2002}.}
\label{fig-hrfm}\end{figure}

The evolutionary tracks are shown on Fig.~\ref{fig-hrfm} for accretion rates of 0.1, 0.01 and 0.001~\Mpy.
Due to the use of various initial models, the early evolution differs between the $\fitm=$~0.999 and 0.99 cases.
But at the stage where the models at 0.001 and 0.1~\Mpy\ converge to their respective asymptotic behaviours,
their evolutionary tracks become nearly independent of \fitm.
The slight shift in the tracks at 0.1~\Mpy\ reflects simply the dependence of the exact location of the Hayashi limit
on the treatment of the external layers of the star, since this limit results from the opacity in the external layers.
The same is true for the 1~\Mpy\ case, not shown here.
In contrast, the asymptotic track in the low-\dm\ regime (model at 0.001~\Mpy) is not affected by a change in \fitm,
because the location of the ZAMS is fixed by the physics in the centre of the star, and a change in the treatment of the outer layers has no impact.
This justifies our choice of $\fitm=0.99$ for the model at 0.001~\Mpy\ described in Sect.~\ref{sec-mod-evol}.

Only for the intermediate case at 0.01 \Mpy, the asymptotic behaviour differs significantly between $\fitm=$~0.999 and 0.99.
After several oscillations in \Teff, when the model at $\fitm=$~0.999 converges to the Hayashi limit in the red, with $\Teff<10^4$ K,
the model at $\fitm=$~0.99 remains in the blue, close to the low-\dm\ tracks, with $\Teff\simeq70\,000$ K.
Thus at this intermediate rate, the asymptotic behaviour is switched from one limit to the other by a change in \fitm.
We notice also that before the convergence to the asymptotic track, the amplitude of the oscillations in \Teff\ is reduced by the decrease in \fitm.

\begin{figure}\begin{center}\includegraphics[width=0.4\textwidth]{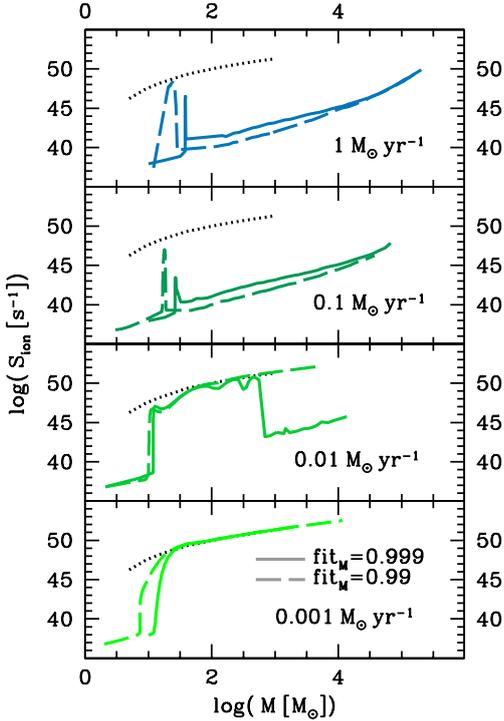}\end{center}
\caption{Ionising flux as a function of the stellar mass for the models with indicated accretion rates.
On each panel, the solid line corresponds to the model with $\fitm=0.999$, and the dashed line to the one with $\fitm=0.99$.
The black dotted line corresponds to the ZAMS of \citealt{schaerer2002}.}
\label{fig-ionfm}\end{figure}

In order to study how this effect impacts the ionising flux, we compute \sion\ according to Eq.~(\ref{eq-ion}) for the same models.
The result is shown on Fig.~\ref{fig-ionfm}.
As expected from the evolutionary tracks, for $\dm\geq0.1$ \Mpy\ and $\dm=0.001$~\Mpy, \sion\ is nearly unaffected by a change in \fitm.
Only in the intermediate case 0.01 \Mpy, \sion\ differs between models at $\fitm=0.999$ and $0.99$.
The reduction of the amplitude of the \Teff-oscillations in the model at $\fitm=0.99$ makes \sion\ to follow the ZAMS limit closer.
But more importantly, as the model with $\fitm=0.999$ converges to the high-\dm\ regime,
the corresponding value of \sion\ decreases suddenly by 8 orders of magnitude, from $10^{51}$ to $10^{43}$\,s$^{-1}$.
At the same stage, the model with $\fitm=0.99$ remains in the blue, with \sion\ growing slowly ($\sion>10^{50}$\,s$^{-1}$).
The 8 orders of magnitude difference is maintained as the stellar mass approaches $10^4$ \Ms.

This example shows that the choice of \fitm\ is critical in order to determine properly the ionising properties
of stars accreting at a rate between 0.001 and 0.1~\Mpy.
Difficulties in numerical convergence make it impossible to use $\fitm>0.999$.
However, the models described above show that an increase in \fitm\ leads to larger radii and lower effective temperatures.
But the Hayashi limit prevents \Teff\ to decrease under 5000 -- 6000 K.
Since our model at $\dm=0.01$ \Mpy\ and $\fitm=0.999$ reaches the Hayashi limit at 600~\Ms,
we do not expect a further increase in \fitm\ to modify the track in the supermassive range.
Actually, regarding the above examples, one can expect convergence to the high-\dm\ asymptotic track to occur earlier at higher \fitm.
Thus a further increase in \fitm\ could potentially reduce the value of \dmc\ closer to 0.001 \Mpy,
and bring definitely the intermediate 0.01~\Mpy\ rate to the high-\dm\ range.
An accurate treatment of the external layers of the accreting star is thus required in order to determine the exact value of \dmc,
but our models suggest a value closer to 0.005~\Mpy\ than to 0.05~\Mpy.

\subsection{Final mass at the onset of the collapse}
\label{sec-dis-mfin}

Despite the absence of hydrodynamics in our models, one can estimate the stage at which the GR instability triggers the collapse
using the polytropic criterion of \cite{chandrasekhar1964} \citep[though see][]{woods2017}.
According to this criterion, the star becomes unstable when the adiabatic index $\Gamma$ is reduced under a critical value $\Gamma_{\rm crit}$.
For a classical star ($GM_r/rc^2=0$), this critical value is simply 4/3.
The first order relativistic development ($GM_r/rc^2<<1$) in the polytropic case gives
\begin{equation}
\Gamma_{\rm crit}={4\over3}+K\,{2GM_r\over r\,c^2}\ ,
\label{eq-gamc1}\end{equation}
where $K$ is a constant that depends on the polytropic index ($K\simeq1.12$ for $n=3$).
For a purely radiation-supported star ($P=P_{\rm rad}$, $P_{\rm gas}=0$), the adiabatic index is exactly 4/3, and the star is unstable.
The first order development in terms of $\beta=P_{\rm gas}/P$, the ratio of gas pressure to total pressure, is
\begin{equation}
\Gamma={4\over3}+{\beta\over6}\ .
\label{eq-gam1}\end{equation}
Thus for stars that are dominated by radiation ($\beta<<1$),
the stability criterion $\Gamma>\Gamma_{\rm crit}$ can be expressed as
\begin{equation}
{\beta\over6}>K\,{2GM_r\over r\,c^2}\ .
\label{eq-crit}\end{equation}

\begin{figure}\begin{center}\includegraphics[width=0.49\textwidth]{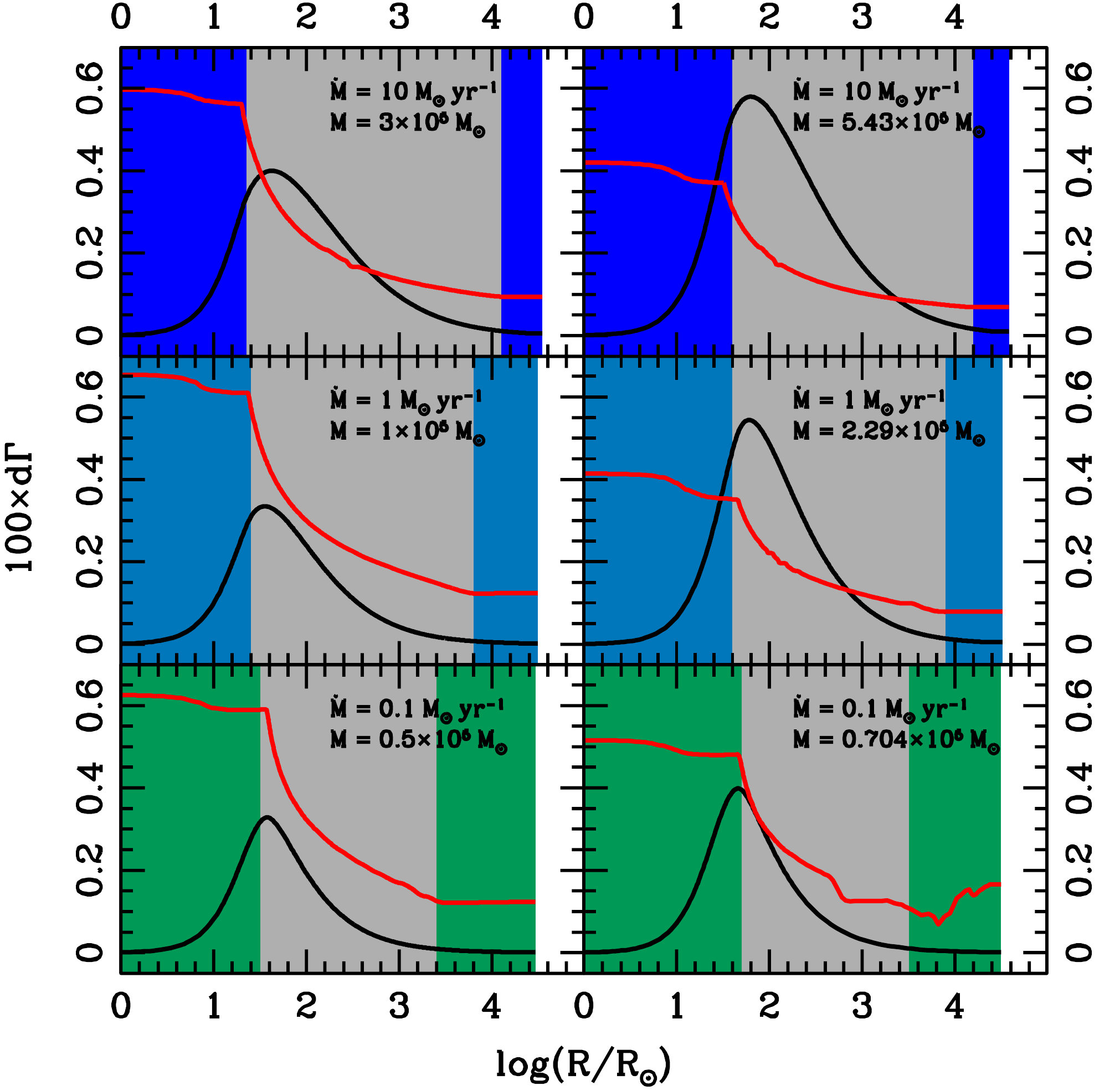}\end{center}
\caption{Polytropic criterion of GR stability for $\dm=10$~\Mpy (upper row), 1~\Mpy (middle row) and 0.1~\Mpy\ (bottom row).
On each plot, the black curves are the internal profiles of the first order GR correction to $\Gamma_{\rm crit}$ (Eq.~\ref{eq-gamc1}) for $K=1.12$ ($n=3$).
The red curves correspond to the internal profiles of the first order correction to $\Gamma$ in terms of $\beta$ (Eq.~\ref{eq-gam1}).
The coloured and grey areas are respectively convective and radiative zones.
The right column shows the profiles at the end of the computation, while the left one shows it a an earlier stage, corresponding to the indicated masses.}
\label{fig-gr}\end{figure}

The two members of Eq.~(\ref{eq-crit}) are compared in Fig.~\ref{fig-gr}
for our models with $\dm=10$, 1 and 0.1 \Mpy\ at various stellar masses, using $K=1.12$.
The right column shows the internal profiles of these quantities at the end of the computations,
while the left one shows it at an earlier stage.
The regions that are unstable according to the polytropic criterion are also shown as red areas on Fig.~\ref{fig-st}.
In all the cases, the stellar structure consist in a convective core (coloured regions at the left-hand sides of each plot),
an intermediate radiative zone (grey regions) and a convective envelope (coloured region at the right-hand sides).
Since the stability criterion of Eq.~(\ref{eq-crit}) is based on polytropic structures,
the changes in the polytropic indices between the various regions of the star complicates the analysis.
In each case, most of the mass of the star is contained in the radiative region, so that one could naively expect the $n=3$ criterion to be relevant.
However, the high-entropy accreted envelope is not approximated by an $n=3$ polytrope in the present case.
Moreover, we notice that the regions that are unstable according to the criterion are located deep in the radiative zone,
close to the edge of the convective core, where the first order relativistic correction to $\Gamma_{\rm crit}$ has its maximum.
In the model at 10~\Mpy, the criterion indicates instability in the radiative zone at $M=261\,000$~\Ms.
As evolution proceeds, the \guig unstable\guid\ zones grow, joining eventually the convective core.
In the model at 1~\Mpy, the criterion indicates instability at $M=116\,000$~\Ms.
The 0.1~\Mpy\ model was stopped at $M=0.7\times10^5\,\Ms$ for numerical reasons,
when $\beta/6$ was approaching $1.12\times2GM_r/rc^2$ close to the convective core.
We summarise these results in Table~\ref{tab-mfin}.
The final masses are compared on Fig.~\ref{fig-mfin} with those of \cite{woods2017} and \cite{umeda2016}, as discussed below (Sect.~\ref{sec-dis-lit}).

\begin{table}\caption{Final masses of the models of indicated accretion rates. The first line is the accretion rate (\dm),
the second one ($M_{\rm crit}$) contains the mass at which instability appears according to the polytropic criterion with index $n=3$,
and the third one (\mfin) shows the final mass of our runs.}
\begin{center}\begin{tabular}{ccccccc}\hline
$\dm=$	&	0.1	&	1	&	10	&	$\Mpy$\\\hline
$M_{\rm crit}=$	&		&	1.16	&	2.61	&	$\times10^5\,\Ms$\\
$\mfin=$	&	0.70	&	2.29	&	5.43	&	$\times10^5\,\Ms$\\\hline
\end{tabular}\end{center}\label{tab-mfin}\end{table}

\subsection{Comparison with previous studies}
\label{sec-dis-lit}

The evolution of Pop III stars under high accretion rates has been computed by several authors under various physical conditions.
\cite{omukai2001,omukai2003} and \cite{hosokawa2012a} used rates in the range 0.001 - 1 \Mpy\ to produce models towards 1000 \Ms\
with the assumption of spherical accretion, in which all the entropy of the accretion shock is advected in the stellar interior.
This assumption is expected to produce stars that are more bloated than in the disc case, due to their high amount of internal entropy.
In these models, stars evolve as red supergiant protostars along the Hayashi line if $\dm>\dmc$, contract to the ZAMS if $\dm<\dmc$,
and oscillate in \Teff\ between these two limits if $\dm\simeq\dmc$.
\cite{omukai2001,omukai2003} obtained a critical value of $\dmc=0.004$~\Mpy.
In \cite{hosokawa2012a}, a model at $\dm=0.006$~\Mpy\ oscillates in \Teff\ without converging to the Hayashi limit
before the end of their computation, at $M=600-700$~\Ms.
\cite{omukai2003} concluded that this expansion indicates the end of the accretion phase,
since radiation pressure close to the Eddington limit is expected to reverse an accretion flow having a spherical symmetry.
In contrast, at lower rates, the star contracts towards the ZAMS despite accretion.
But at the high effective temperatures of the ZAMS, the ionising effect of the radiation field becomes significant,
and a large \hii\ region, with high pressure, is expected to form around the star, preventing accretion above $\sim50$ \Ms\ \citep{hosokawa2011b}.

All these computations were stopped at 1000 \Ms\ due to convergence difficulties in the code used,
that solves the structure equations with a shooting method.
\cite{hosokawa2013} used instead the numerical code {\sc stellar} \citep{yorke2008}, based on the Henyey method, like the \gva\ code,
to push the computations towards higher masses.
In contrast to the previous studies, entropy is accreted according to cold disc accretion, like in our models.
They confirmed that at rates above \dmc\ the star evolves as a red supergiant protostar, with \Teff\ $<10^4$ K.
But the value they obtain for \dmc\ exceeds 0.01~\Mpy, in contrast to \cite{omukai2001,omukai2003}.
Indeed, after several oscillations in \Teff, their model at 0.01~\Mpy\ remains in the blue until the end of their computation,
at $M\simeq2000$~\Ms\ and $L\simeq10^8$~\Ls.
Only at 0.1~\Mpy\ the star converges to the Hayashi limit, already at $M\simeq30$~\Ms, and evolves then as a red supergiant protostar.
Exploring the supermassive range, the supergiant models of \cite{hosokawa2013} start to decrease in radius at $M\simeq3\times10^4$ \Ms.
At this point, the effective temperature grows,
which could potentially lead to the formation of an \hii\ region during further evolution, not followed in their models.
Their computation stops at $10^5$~\Ms.
Although their code does not include the GR correction in general, they computed a few test-models with it,
and found the correction to be negligible below $10^5$ \Ms, when they stopped their computations.
The difference in the exact value of \dmc\ between \cite{omukai2001,omukai2003} and \cite{hosokawa2013}
is expected to come from the change in the accretion of entropy, between hot spherical and cold disc accretion.
Indeed, the reduction of the entropy accreted in the disc case compared to the spherical case makes accreting stars more compact,
so that a higher rate is needed in order to produce red supergiant protostars.
Thus the results of \cite{omukai2001,omukai2003} and \cite{hosokawa2013}
indicate $\dmc\simeq0.004$~\Mpy\ for spherical accretion and $\dmc>0.01$~\Mpy\ for disc accretion.

Notice that \cite{schleicher2013}, using analytical considerations based on timescales comparisons,
obtained that stars accreting at a rate above 0.14~\Mpy\ continue to expand in radius until the mass reaches $10^6$~\Ms,
in contrast with the results of \cite{hosokawa2013}.

Our models, based on cold disc accretion, explore the highest mass-range, above $10^5$~\Ms, until the star collapse into a black hole.
We confirm the main features obtained in the studies mentioned above:
stars accreting above a critical rate \dmc\ evolve as red supergiants along the Hayashi limit, while under \dmc\ they contract towards the ZAMS.
However, two differences appear between our models and those of \cite{hosokawa2013}.
First, our results indicate $\dmc<0.01$~\Mpy, since after several oscillations in effective temperature
our model at 0.01~\Mpy\ converges to the Hayashi limit and evolve as a red supergiant protostar when its mass exceeds 600~\Ms.
This fact suggests that \dmc\ is closer to the value obtained by \cite{omukai2001,omukai2003} for spherical accretion
than to the one obtained by \cite{hosokawa2013} in the disc case.
Our numerical experiment described in Sect.~\ref{sec-dis-fm} suggests that this discrepancy is related
to the treatment of the energy equation in the external layers of the star.
Indeed, neglecting the production of energy on the external 1\% of the stellar mass
gives an evolutionary track at 0.01~\Mpy\ which is similar as that of \cite{hosokawa2013}.
In other words, our models show that an accurate treatment of the energy equation in the external layers
can reduce the value of \dmc\ by nearly one order of magnitude, from $\sim0.05$ to $\sim0.005$~\Mpy,
extending the range of accretion rate at which the ionising feedback remains negligible.

The second difference between our models and those of \cite{hosokawa2013} concerns the evolution in the supermassive range.
In contrast to the models of \cite{hosokawa2013}, our supergiant models never show a decrease of the stellar radius at $M>10^4$~\Ms.
Until the highest mass ($\simeq5\times10^5$~\Ms\ for $\dm=10$~\Mpy), the radius continues to expand and \Teff\ remains $\simeq10^4$ K.
Thus our models confirm the analytical results of \cite{schleicher2013}, in contrast to \cite{hosokawa2013},
and we do not expect the ionising flux to grow significantly at $M>10^5$~\Ms.
This result extends the mass-range in which stars accreting at high rate do not significantly ionise their surrounding.

\quad

Recently, \cite{umeda2016} presented models of Pop~III stars accreting at rates 0.1 -- 0.3 -- 1 -- 10~\Mpy, including the post-Newtonian correction.
Their model run until final masses of 1.2 -- 1.9 --  3.5 -- $8.0\times10^5$~\Ms, respectively.
At these masses, the polytropic criterion of Eq.~(\ref{eq-crit}) with $n=3$ indicates instability in the convective core.
Actually, the criterion indicates instability in the core already {\it before} the star reaches these final masses, as it is visible on their Fig.~3.
No evolutionary track of their models are provided, and we do not know if expansion stops in their model when the stellar mass exceeds $10^4$~\Ms.

In a previous paper \citep{woods2017}, we used the \kep\ code to follow the evolution of Pop III stars accreting at high rates until the final collapse.
Since the \kep\ code includes hydrodynamics, the onset of the collapse can be followed self-consistently,
without the use of the polytropic criterion considered in the present work and in \cite{umeda2016}.
The collapse is triggered at masses of $1.5-3.3\times10^5$~\Ms\ for $\dm=0.1-10$~\Mpy.

\begin{figure}\begin{center}\includegraphics[width=0.49\textwidth]{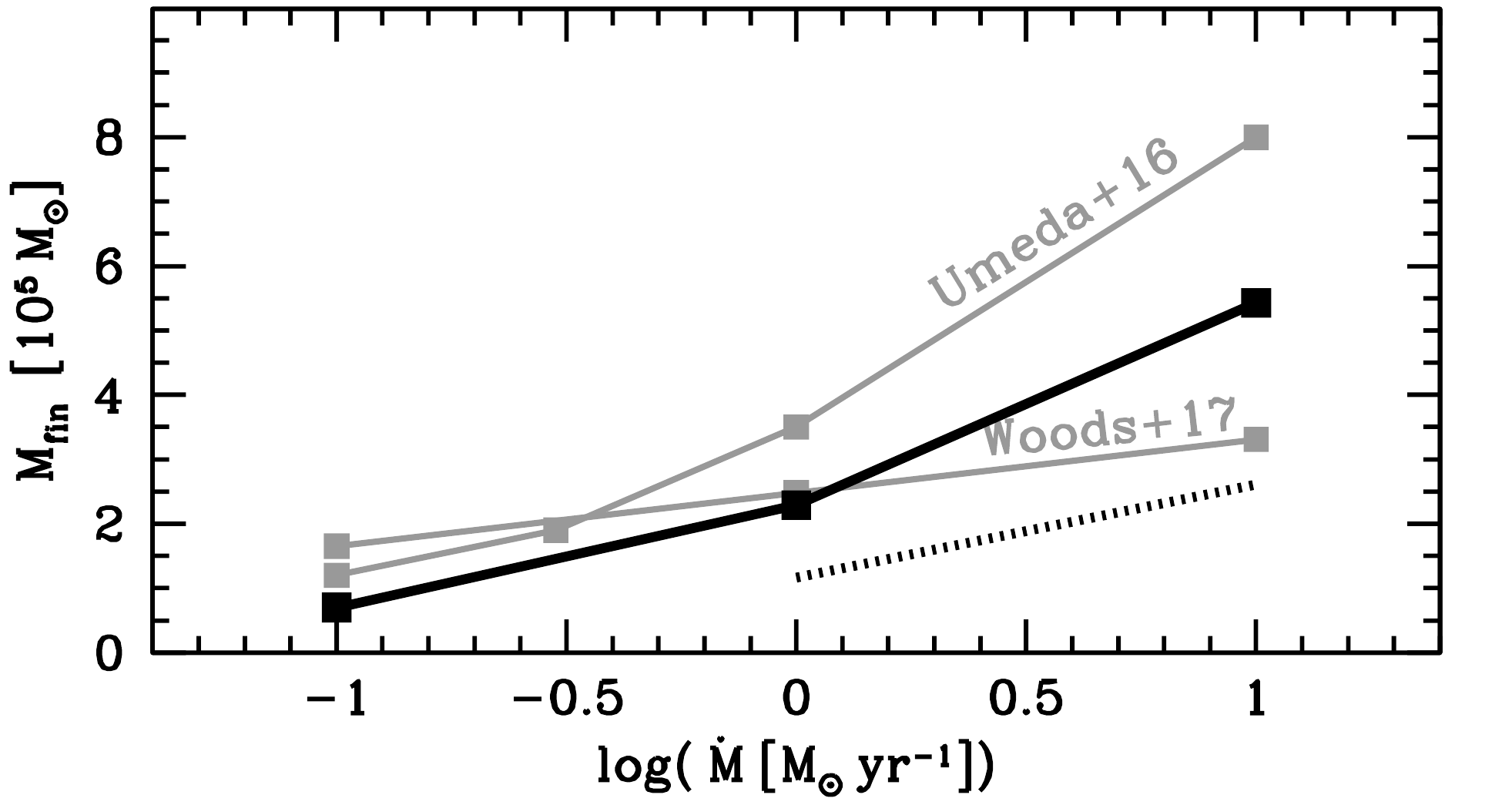}\end{center}
\caption{Final masses as a function of the accretion rate.
The black curves show the results of the present work: the solid line corresponds the final mass of our runs,
and the dotted line to the mass at which the polytropic criterion indicates instability for $n=3$.
The grey curves are the final masses according to previous studies \citep{woods2017,umeda2016}.}
\label{fig-mfin}\end{figure}

The final masses are shown on Fig.~\ref{fig-mfin} as a function of the accretion rate, for the various studies.
Models of \cite{umeda2016}, of \cite{woods2017} and of the present work all agree with the fact that the mass at collapse
remains in the same order of magnitude ($\sim1-8\times10^5$~\Ms) over the two orders of magnitude range from 0.1 to 10~\Mpy\ in \dm,
with a slight increase as a function of \dm.
However, strong discrepancies appear in the exact values of the mass.
\cite{umeda2016} obtained final masses that are larger to our \mfin\ by a factor 1.5, nearly independently of the accretion rate.
Thus, despite the discrepancies in \mfin, the dependence of \mfin\ on \dm\ is similar between \cite{umeda2016} and us.
In \cite{woods2017}, the slope of the curve (\dm,\,\mfin), i.e. the dependence of \mfin\ on \dm,
is weaker than in the present models and those of \cite{umeda2016}.
In particular, for $\dm=10$~\Mpy, the final mass in \cite{woods2017} is only 60\% of that of the present work.

The validity of the polytropic criterion used in the present work can be questioned,
since the stellar structures considered here are not polytropes.
Indeed, \cite{woods2017} showed that with a self-consistent treatment of hydrodynamics
the star remains stable well after the polytropic criterion indicates instability in the convective core.
This suggests that the criterion provides only a lower limit for the final mass of SMSs.

Moreover, the final masses of our runs are not necessary the masses at collapse,
first because numerical instability could be responsible for the impossibility to make the code to converge at higher masses,
second because the GR instability is a pulsational instability \citep{chandrasekhar1964}.
Marginal stability allows hydrostatic equations to be solved during that stage, and a hydrostatic code can evolve through this phase without noticing anything.
As a consequence, a fully consistent treatment of hydrodynamics is necessary to capture the GR instability.

A detailed treatment of the accretion of entropy and of convection itself are expected significantly effect the results.
Any departure from cold disc accretion could impact the stellar structure, in particular at high rates,
for which the accretion history of entropy plays a more important role than internal entropy redistribution in shaping the entropy gradient,
as described in Sect.~\ref{sec-mod-evol}.
Since the entropy gradient determines the presence of convection,
a change in the accretion of entropy could impact significantly the mass at which the collapse occurs.
More importantly, the treatment of convection, based on the mixing-length theory, appears as critical.
In the present work, the triggering of convection is based on the Schwarzschild criterion.
The use of the Ledoux criterion, that takes the chemical gradient into account as a stabilizing effect,
favours purely radiative transport and reduces convective regions.
Moreover, we do not include overshooting in our models.
Including overshooting would increase the mass of the convective core, and thus decrease potentially the collapse masses.
As a consequence, a precise estimation of the stellar mass at collapse is currently problematic,
and one can interpret the various curves of Fig.~\ref{fig-mfin} as providing the envelope of the real $\dm-\mfin$ relation.

\section{Conclusion}
\label{sec-out}

In the present work, we described new models of Pop III protostars accreting at high rates (0.001 -- 10~\Mpy)
towards the mass at which the GR instability is expected to trigger the collapse into a black hole.
We described the evolutionary tracks and internal structures of these models, and studied the properties of their radiative feedback.
We confirm the results of previous studies that stars accreting at a rate above a critical value \dmc\ evolve as red supergiants,
following the Hayashi limit in the red part of the HR diagram with a weak ionising feedback,
while for lower rates the star contracts towards the ZAMS in the blue with a strong ionising feedback.

In contrast to previous studies, our models show that Pop III protostars accreting at rates $\gtrsim0.1$~\Mpy\
continue to expand in radius in the highest mass-range ($>10^5$~\Ms), until it reaches GR instability.
Thus no significant increase in the ionising effect of the radiation field are expected in this mass-range as long as rapid accretion proceeds.
In addition, compared to previous studies, our models reduce the value of \dmc\ in the case of cold disc accretion,
by nearly one order of magnitude, from $\sim0.05$ to $\sim0.005$~\Mpy.
Thus our results extend the range of masses and accretion rates
at which the ionising feedback remains negligible.

Using the polytropic criterion of \cite{chandrasekhar1964} for GR instability, we estimated the mass at which the protostar collapses into a black hole.
We obtained collapse masses that remain in the same order of magnitude (ranging from 0.7 to $5.5\times10^5$~\Ms)
for accretion rates that varies in the two orders of magnitude range 0.1 -- 10~\Mpy.
Inside this interval, the collapse mass increases with \dm.
Our final masses are in the interval of the various values obtained in previous studies,
and the dependence on \dm\ is qualitatively in agreement with these studies.
Discrepancies remain in the exact value of the mass at collapse.
We interpret them as the result of the various treatment of entropy accretion and of convection,
and in particular the use of the polytropic criterion instead of a fully consistent treatment of the hydrodynamics.

\appendix

\section{Tables}

Tables~\ref{tab-1ep1} to \ref{tab-1em3} display the age, mass, radius, luminosity, effective temperature, ionising flux and Lyman-Werner flux.
The ionising and LW fluxes are computed from Eq.~(\ref{eq-ion}) and (\ref{eq-lw}).
The age is counted since $M=0$, i.e. $t=M/\dot M$.
The time-steps are computed so that the differences in $\log(L/\Ls)$ and $\log\Teff$ between two consecutive points
does not exceed 0.1 and 0.025, respectively.

\begin{table*}\caption{Model at $\dm=10$ \Mpy}\begin{center}\begin{tabular}{|c|c|c|c|c|c|c|}\hline
age [yr] & log($M$/\Ms) & log($R$/\Rs) & log($L$/\Ls) & log(\Teff[K]) & log(\sion[s$^{-1}$]) & log(\slw[s$^{-1}$]) \\\hline
1.000e+00 & 1.000e+00 & 2.235e+00 & 4.133e+00 & 3.678e+00 & 3.794e+01 & 4.022e+01 \\
3.376e+02 & 1.348e+00 & 2.187e+00 & 4.137e+00 & 3.703e+00 & 3.857e+01 & 4.081e+01 \\
3.394e+02 & 1.602e+00 & 2.070e+00 & 4.002e+00 & 3.728e+00 & 3.912e+01 & 4.120e+01 \\
3.403e+02 & 1.693e+00 & 2.029e+00 & 4.023e+00 & 3.753e+00 & 3.984e+01 & 4.175e+01 \\
3.403e+02 & 1.693e+00 & 2.035e+00 & 4.127e+00 & 3.776e+00 & 4.048e+01 & 4.233e+01 \\
3.403e+02 & 1.694e+00 & 2.038e+00 & 4.230e+00 & 3.800e+00 & 4.113e+01 & 4.285e+01 \\
3.403e+02 & 1.695e+00 & 2.042e+00 & 4.330e+00 & 3.824e+00 & 4.173e+01 & 4.336e+01 \\
3.403e+02 & 1.695e+00 & 2.046e+00 & 4.434e+00 & 3.847e+00 & 4.227e+01 & 4.383e+01 \\
3.403e+02 & 1.696e+00 & 2.051e+00 & 4.540e+00 & 3.872e+00 & 4.282e+01 & 4.430e+01 \\
3.403e+02 & 1.696e+00 & 2.056e+00 & 4.641e+00 & 3.894e+00 & 4.332e+01 & 4.469e+01 \\
3.403e+02 & 1.697e+00 & 2.062e+00 & 4.742e+00 & 3.917e+00 & 4.383e+01 & 4.508e+01 \\
3.403e+02 & 1.697e+00 & 2.069e+00 & 4.846e+00 & 3.939e+00 & 4.425e+01 & 4.547e+01 \\
3.403e+02 & 1.698e+00 & 2.076e+00 & 4.953e+00 & 3.962e+00 & 4.471e+01 & 4.582e+01 \\
3.404e+02 & 1.698e+00 & 2.085e+00 & 5.059e+00 & 3.984e+00 & 4.510e+01 & 4.617e+01 \\
3.404e+02 & 1.699e+00 & 2.096e+00 & 5.162e+00 & 4.005e+00 & 4.549e+01 & 4.647e+01 \\
3.404e+02 & 1.700e+00 & 2.109e+00 & 5.267e+00 & 4.024e+00 & 4.583e+01 & 4.677e+01 \\
3.404e+02 & 1.700e+00 & 2.126e+00 & 5.369e+00 & 4.041e+00 & 4.614e+01 & 4.702e+01 \\
3.404e+02 & 1.701e+00 & 2.148e+00 & 5.470e+00 & 4.056e+00 & 4.639e+01 & 4.724e+01 \\
3.404e+02 & 1.702e+00 & 2.179e+00 & 5.573e+00 & 4.065e+00 & 4.660e+01 & 4.742e+01 \\
3.404e+02 & 1.704e+00 & 2.228e+00 & 5.675e+00 & 4.067e+00 & 4.672e+01 & 4.753e+01 \\
3.404e+02 & 1.706e+00 & 2.309e+00 & 5.775e+00 & 4.051e+00 & 4.663e+01 & 4.751e+01 \\
3.405e+02 & 1.709e+00 & 2.496e+00 & 5.875e+00 & 3.983e+00 & 4.592e+01 & 4.697e+01 \\
3.406e+02 & 1.716e+00 & 3.005e+00 & 5.980e+00 & 3.754e+00 & 4.180e+01 & 4.378e+01 \\
3.406e+02 & 1.717e+00 & 3.096e+00 & 6.081e+00 & 3.734e+00 & 4.141e+01 & 4.342e+01 \\
3.406e+02 & 1.718e+00 & 3.165e+00 & 6.184e+00 & 3.726e+00 & 4.130e+01 & 4.339e+01 \\
3.406e+02 & 1.719e+00 & 3.218e+00 & 6.284e+00 & 3.724e+00 & 4.141e+01 & 4.341e+01 \\
3.406e+02 & 1.720e+00 & 3.271e+00 & 6.388e+00 & 3.724e+00 & 4.140e+01 & 4.352e+01 \\
3.406e+02 & 1.722e+00 & 3.323e+00 & 6.492e+00 & 3.724e+00 & 4.150e+01 & 4.362e+01 \\
3.407e+02 & 1.723e+00 & 3.371e+00 & 6.592e+00 & 3.725e+00 & 4.171e+01 & 4.380e+01 \\
3.407e+02 & 1.725e+00 & 3.440e+00 & 6.768e+00 & 3.734e+00 & 4.210e+01 & 4.411e+01 \\
3.407e+02 & 1.729e+00 & 3.487e+00 & 6.869e+00 & 3.736e+00 & 4.220e+01 & 4.428e+01 \\
3.408e+02 & 1.732e+00 & 3.537e+00 & 6.990e+00 & 3.741e+00 & 4.252e+01 & 4.453e+01 \\
3.928e+02 & 2.759e+00 & 3.571e+00 & 7.090e+00 & 3.749e+00 & 4.282e+01 & 4.476e+01 \\
4.041e+02 & 2.837e+00 & 3.617e+00 & 7.190e+00 & 3.751e+00 & 4.291e+01 & 4.492e+01 \\
4.182e+02 & 2.918e+00 & 3.663e+00 & 7.290e+00 & 3.753e+00 & 4.311e+01 & 4.502e+01 \\
4.353e+02 & 3.000e+00 & 3.710e+00 & 7.390e+00 & 3.755e+00 & 4.321e+01 & 4.519e+01 \\
4.566e+02 & 3.084e+00 & 3.754e+00 & 7.490e+00 & 3.757e+00 & 4.340e+01 & 4.535e+01 \\
4.828e+02 & 3.169e+00 & 3.800e+00 & 7.591e+00 & 3.760e+00 & 4.350e+01 & 4.544e+01 \\
5.153e+02 & 3.255e+00 & 3.846e+00 & 7.691e+00 & 3.761e+00 & 4.369e+01 & 4.561e+01 \\
5.556e+02 & 3.343e+00 & 3.892e+00 & 7.791e+00 & 3.764e+00 & 4.388e+01 & 4.577e+01 \\
6.054e+02 & 3.431e+00 & 3.938e+00 & 7.891e+00 & 3.765e+00 & 4.398e+01 & 4.586e+01 \\
6.675e+02 & 3.521e+00 & 3.983e+00 & 7.991e+00 & 3.768e+00 & 4.417e+01 & 4.602e+01 \\
7.444e+02 & 3.612e+00 & 4.027e+00 & 8.091e+00 & 3.771e+00 & 4.436e+01 & 4.618e+01 \\
8.394e+02 & 3.702e+00 & 4.072e+00 & 8.191e+00 & 3.774e+00 & 4.446e+01 & 4.634e+01 \\
9.581e+02 & 3.794e+00 & 4.117e+00 & 8.291e+00 & 3.776e+00 & 4.464e+01 & 4.649e+01 \\
1.107e+03 & 3.887e+00 & 4.159e+00 & 8.391e+00 & 3.780e+00 & 4.482e+01 & 4.665e+01 \\
1.292e+03 & 3.981e+00 & 4.201e+00 & 8.491e+00 & 3.784e+00 & 4.501e+01 & 4.685e+01 \\
1.524e+03 & 4.075e+00 & 4.242e+00 & 8.591e+00 & 3.789e+00 & 4.519e+01 & 4.701e+01 \\
1.813e+03 & 4.170e+00 & 4.284e+00 & 8.691e+00 & 3.793e+00 & 4.544e+01 & 4.721e+01 \\
2.177e+03 & 4.265e+00 & 4.322e+00 & 8.791e+00 & 3.799e+00 & 4.562e+01 & 4.741e+01 \\
2.632e+03 & 4.361e+00 & 4.361e+00 & 8.891e+00 & 3.804e+00 & 4.587e+01 & 4.761e+01 \\
3.200e+03 & 4.457e+00 & 4.397e+00 & 8.991e+00 & 3.811e+00 & 4.611e+01 & 4.780e+01 \\
3.912e+03 & 4.554e+00 & 4.432e+00 & 9.091e+00 & 3.819e+00 & 4.635e+01 & 4.804e+01 \\
4.800e+03 & 4.650e+00 & 4.466e+00 & 9.191e+00 & 3.827e+00 & 4.665e+01 & 4.827e+01 \\
5.918e+03 & 4.747e+00 & 4.494e+00 & 9.291e+00 & 3.838e+00 & 4.694e+01 & 4.853e+01 \\
7.328e+03 & 4.845e+00 & 4.512e+00 & 9.391e+00 & 3.854e+00 & 4.734e+01 & 4.887e+01 \\
9.081e+03 & 4.942e+00 & 4.520e+00 & 9.491e+00 & 3.875e+00 & 4.783e+01 & 4.929e+01 \\
1.020e+04 & 4.994e+00 & 4.496e+00 & 9.544e+00 & 3.900e+00 & 4.832e+01 & 4.969e+01 \\
1.191e+04 & 5.064e+00 & 4.482e+00 & 9.616e+00 & 3.925e+00 & 4.882e+01 & 5.006e+01 \\
1.486e+04 & 5.162e+00 & 4.533e+00 & 9.716e+00 & 3.924e+00 & 4.892e+01 & 5.016e+01 \\
1.855e+04 & 5.260e+00 & 4.537e+00 & 9.816e+00 & 3.947e+00 & 4.937e+01 & 5.053e+01 \\
2.319e+04 & 5.359e+00 & 4.538e+00 & 9.916e+00 & 3.972e+00 & 4.980e+01 & 5.091e+01 \\
2.901e+04 & 5.458e+00 & 4.543e+00 & 1.002e+01 & 3.994e+00 & 5.021e+01 & 5.124e+01 \\
\end{tabular}\end{center}\label{tab-1ep1}\end{table*}
\begin{table*}\contcaption{}\begin{center}\begin{tabular}{|c|c|c|c|c|c|c|}
3.627e+04 & 5.556e+00 & 4.556e+00 & 1.012e+01 & 4.013e+00 & 5.055e+01 & 5.151e+01 \\
4.540e+04 & 5.654e+00 & 4.567e+00 & 1.022e+01 & 4.033e+00 & 5.087e+01 & 5.179e+01 \\
\hline\end{tabular}\end{center}\label{tab-1ep1}\end{table*}

\begin{table*}\caption{Model at $\dm=1$ \Mpy}\begin{center}\begin{tabular}{|c|c|c|c|c|c|c|}\hline
age [yr] & log($M$/\Ms) & log($R$/\Rs) & log($L$/\Ls) & log(\Teff[K]) & log(\sion[s$^{-1}$]) & log(\slw[s$^{-1}$]) \\\hline
1.000e+01 & 1.000e+00 & 2.235e+00 & 4.133e+00 & 3.678e+00 & 3.794e+01 & 4.022e+01 \\
3.540e+02 & 1.270e+00 & 2.098e+00 & 3.958e+00 & 3.703e+00 & 3.840e+01 & 4.064e+01 \\
3.673e+02 & 1.504e+00 & 1.965e+00 & 3.793e+00 & 3.728e+00 & 3.891e+01 & 4.099e+01 \\
3.732e+02 & 1.578e+00 & 1.975e+00 & 3.894e+00 & 3.748e+00 & 3.952e+01 & 4.157e+01 \\
3.732e+02 & 1.578e+00 & 1.987e+00 & 4.003e+00 & 3.769e+00 & 4.018e+01 & 4.209e+01 \\
3.733e+02 & 1.579e+00 & 1.992e+00 & 4.109e+00 & 3.793e+00 & 4.086e+01 & 4.263e+01 \\
3.733e+02 & 1.579e+00 & 1.996e+00 & 4.214e+00 & 3.817e+00 & 4.147e+01 & 4.311e+01 \\
3.733e+02 & 1.579e+00 & 2.001e+00 & 4.318e+00 & 3.841e+00 & 4.203e+01 & 4.360e+01 \\
3.733e+02 & 1.579e+00 & 2.006e+00 & 4.423e+00 & 3.865e+00 & 4.260e+01 & 4.408e+01 \\
3.733e+02 & 1.580e+00 & 2.012e+00 & 4.528e+00 & 3.888e+00 & 4.312e+01 & 4.448e+01 \\
3.734e+02 & 1.580e+00 & 2.019e+00 & 4.631e+00 & 3.910e+00 & 4.359e+01 & 4.488e+01 \\
3.734e+02 & 1.580e+00 & 2.026e+00 & 4.734e+00 & 3.932e+00 & 4.406e+01 & 4.528e+01 \\
3.734e+02 & 1.580e+00 & 2.034e+00 & 4.842e+00 & 3.955e+00 & 4.450e+01 & 4.565e+01 \\
3.734e+02 & 1.581e+00 & 2.044e+00 & 4.950e+00 & 3.977e+00 & 4.490e+01 & 4.601e+01 \\
3.735e+02 & 1.581e+00 & 2.055e+00 & 5.053e+00 & 3.998e+00 & 4.527e+01 & 4.629e+01 \\
3.735e+02 & 1.581e+00 & 2.069e+00 & 5.156e+00 & 4.017e+00 & 4.562e+01 & 4.658e+01 \\
3.735e+02 & 1.582e+00 & 2.086e+00 & 5.260e+00 & 4.034e+00 & 4.594e+01 & 4.683e+01 \\
3.736e+02 & 1.582e+00 & 2.110e+00 & 5.363e+00 & 4.048e+00 & 4.620e+01 & 4.707e+01 \\
3.736e+02 & 1.583e+00 & 2.142e+00 & 5.465e+00 & 4.057e+00 & 4.641e+01 & 4.725e+01 \\
3.737e+02 & 1.583e+00 & 2.190e+00 & 5.566e+00 & 4.058e+00 & 4.651e+01 & 4.736e+01 \\
3.737e+02 & 1.584e+00 & 2.269e+00 & 5.666e+00 & 4.044e+00 & 4.646e+01 & 4.734e+01 \\
3.738e+02 & 1.585e+00 & 2.437e+00 & 5.767e+00 & 3.985e+00 & 4.584e+01 & 4.690e+01 \\
3.742e+02 & 1.589e+00 & 2.954e+00 & 5.876e+00 & 3.754e+00 & 4.170e+01 & 4.367e+01 \\
3.742e+02 & 1.589e+00 & 3.050e+00 & 5.989e+00 & 3.734e+00 & 4.132e+01 & 4.333e+01 \\
3.743e+02 & 1.590e+00 & 3.116e+00 & 6.097e+00 & 3.728e+00 & 4.121e+01 & 4.337e+01 \\
3.750e+02 & 1.598e+00 & 3.186e+00 & 6.197e+00 & 3.718e+00 & 4.110e+01 & 4.326e+01 \\
4.830e+02 & 2.169e+00 & 3.214e+00 & 6.298e+00 & 3.729e+00 & 4.152e+01 & 4.357e+01 \\
5.024e+02 & 2.223e+00 & 3.278e+00 & 6.398e+00 & 3.722e+00 & 4.141e+01 & 4.353e+01 \\
5.267e+02 & 2.282e+00 & 3.324e+00 & 6.498e+00 & 3.724e+00 & 4.162e+01 & 4.363e+01 \\
5.607e+02 & 2.353e+00 & 3.366e+00 & 6.598e+00 & 3.729e+00 & 4.183e+01 & 4.387e+01 \\
6.024e+02 & 2.427e+00 & 3.409e+00 & 6.698e+00 & 3.732e+00 & 4.192e+01 & 4.404e+01 \\
6.519e+02 & 2.500e+00 & 3.453e+00 & 6.798e+00 & 3.735e+00 & 4.213e+01 & 4.421e+01 \\
7.105e+02 & 2.574e+00 & 3.496e+00 & 6.898e+00 & 3.739e+00 & 4.233e+01 & 4.438e+01 \\
7.846e+02 & 2.653e+00 & 3.541e+00 & 6.999e+00 & 3.741e+00 & 4.253e+01 & 4.454e+01 \\
8.727e+02 & 2.730e+00 & 3.586e+00 & 7.099e+00 & 3.744e+00 & 4.263e+01 & 4.464e+01 \\
9.819e+02 & 2.811e+00 & 3.630e+00 & 7.199e+00 & 3.747e+00 & 4.283e+01 & 4.480e+01 \\
1.117e+03 & 2.893e+00 & 3.675e+00 & 7.299e+00 & 3.749e+00 & 4.302e+01 & 4.497e+01 \\
1.286e+03 & 2.978e+00 & 3.720e+00 & 7.399e+00 & 3.752e+00 & 4.312e+01 & 4.513e+01 \\
1.495e+03 & 3.064e+00 & 3.766e+00 & 7.499e+00 & 3.754e+00 & 4.332e+01 & 4.529e+01 \\
1.754e+03 & 3.152e+00 & 3.811e+00 & 7.599e+00 & 3.756e+00 & 4.351e+01 & 4.539e+01 \\
2.075e+03 & 3.241e+00 & 3.855e+00 & 7.699e+00 & 3.759e+00 & 4.361e+01 & 4.555e+01 \\
2.475e+03 & 3.330e+00 & 3.900e+00 & 7.799e+00 & 3.762e+00 & 4.380e+01 & 4.571e+01 \\
2.969e+03 & 3.421e+00 & 3.945e+00 & 7.899e+00 & 3.764e+00 & 4.399e+01 & 4.587e+01 \\
3.584e+03 & 3.512e+00 & 3.989e+00 & 7.999e+00 & 3.767e+00 & 4.418e+01 & 4.603e+01 \\
4.350e+03 & 3.604e+00 & 4.032e+00 & 8.099e+00 & 3.771e+00 & 4.437e+01 & 4.619e+01 \\
5.308e+03 & 3.697e+00 & 4.075e+00 & 8.199e+00 & 3.774e+00 & 4.446e+01 & 4.634e+01 \\
6.500e+03 & 3.790e+00 & 4.117e+00 & 8.299e+00 & 3.778e+00 & 4.465e+01 & 4.650e+01 \\
8.002e+03 & 3.885e+00 & 4.158e+00 & 8.399e+00 & 3.783e+00 & 4.492e+01 & 4.671e+01 \\
9.863e+03 & 3.979e+00 & 4.199e+00 & 8.499e+00 & 3.787e+00 & 4.509e+01 & 4.692e+01 \\
1.220e+04 & 4.074e+00 & 4.239e+00 & 8.599e+00 & 3.792e+00 & 4.527e+01 & 4.706e+01 \\
1.512e+04 & 4.170e+00 & 4.278e+00 & 8.699e+00 & 3.798e+00 & 4.553e+01 & 4.727e+01 \\
1.877e+04 & 4.266e+00 & 4.316e+00 & 8.799e+00 & 3.804e+00 & 4.578e+01 & 4.746e+01 \\
2.336e+04 & 4.362e+00 & 4.351e+00 & 8.899e+00 & 3.811e+00 & 4.602e+01 & 4.771e+01 \\
2.907e+04 & 4.458e+00 & 4.386e+00 & 8.999e+00 & 3.819e+00 & 4.626e+01 & 4.794e+01 \\
3.623e+04 & 4.555e+00 & 4.418e+00 & 9.099e+00 & 3.828e+00 & 4.656e+01 & 4.817e+01 \\
4.520e+04 & 4.652e+00 & 4.444e+00 & 9.199e+00 & 3.840e+00 & 4.691e+01 & 4.848e+01 \\
5.644e+04 & 4.749e+00 & 4.463e+00 & 9.299e+00 & 3.855e+00 & 4.731e+01 & 4.881e+01 \\
7.047e+04 & 4.846e+00 & 4.481e+00 & 9.399e+00 & 3.871e+00 & 4.768e+01 & 4.912e+01 \\
8.815e+04 & 4.944e+00 & 4.493e+00 & 9.499e+00 & 3.891e+00 & 4.814e+01 & 4.952e+01 \\
1.089e+05 & 5.036e+00 & 4.490e+00 & 9.593e+00 & 3.916e+00 & 4.863e+01 & 4.993e+01 \\
1.363e+05 & 5.133e+00 & 4.511e+00 & 9.693e+00 & 3.930e+00 & 4.898e+01 & 5.022e+01 \\
1.672e+05 & 5.222e+00 & 4.506e+00 & 9.783e+00 & 3.955e+00 & 4.944e+01 & 5.059e+01 \\
2.052e+05 & 5.311e+00 & 4.501e+00 & 9.874e+00 & 3.980e+00 & 4.989e+01 & 5.095e+01 \\
\hline\end{tabular}\end{center}\label{tab-1e00}\end{table*}

\begin{table*}\caption{Model at $\dm=0.1$ \Mpy}\begin{center}\begin{tabular}{|c|c|c|c|c|c|c|}\hline
age [yr] & log($M$/\Ms) & log($R$/\Rs) & log($L$/\Ls) & log(\Teff[K]) & log(\sion[s$^{-1}$]) & log(\slw[s$^{-1}$]) \\\hline
1.000e+02 & 1.000e+00 & 2.235e+00 & 4.133e+00 & 3.678e+00 & 3.794e+01 & 4.022e+01 \\
5.172e+02 & 1.260e+00 & 2.090e+00 & 3.943e+00 & 3.703e+00 & 3.838e+01 & 4.062e+01 \\
5.936e+02 & 1.412e+00 & 2.112e+00 & 4.043e+00 & 3.717e+00 & 3.895e+01 & 4.103e+01 \\
5.975e+02 & 1.419e+00 & 2.155e+00 & 4.144e+00 & 3.721e+00 & 3.916e+01 & 4.120e+01 \\
5.982e+02 & 1.420e+00 & 2.185e+00 & 4.254e+00 & 3.733e+00 & 3.959e+01 & 4.160e+01 \\
5.983e+02 & 1.420e+00 & 2.199e+00 & 4.360e+00 & 3.752e+00 & 4.018e+01 & 4.209e+01 \\
5.985e+02 & 1.420e+00 & 2.210e+00 & 4.463e+00 & 3.773e+00 & 4.073e+01 & 4.261e+01 \\
5.987e+02 & 1.421e+00 & 2.222e+00 & 4.567e+00 & 3.793e+00 & 4.132e+01 & 4.303e+01 \\
5.990e+02 & 1.421e+00 & 2.236e+00 & 4.669e+00 & 3.811e+00 & 4.179e+01 & 4.348e+01 \\
5.993e+02 & 1.422e+00 & 2.253e+00 & 4.770e+00 & 3.828e+00 & 4.223e+01 & 4.384e+01 \\
5.998e+02 & 1.422e+00 & 2.273e+00 & 4.872e+00 & 3.843e+00 & 4.265e+01 & 4.419e+01 \\
6.003e+02 & 1.423e+00 & 2.298e+00 & 4.974e+00 & 3.856e+00 & 4.298e+01 & 4.448e+01 \\
6.011e+02 & 1.424e+00 & 2.331e+00 & 5.076e+00 & 3.865e+00 & 4.325e+01 & 4.473e+01 \\
6.022e+02 & 1.426e+00 & 2.373e+00 & 5.176e+00 & 3.869e+00 & 4.346e+01 & 4.490e+01 \\
6.038e+02 & 1.429e+00 & 2.431e+00 & 5.277e+00 & 3.866e+00 & 4.345e+01 & 4.493e+01 \\
6.059e+02 & 1.432e+00 & 2.512e+00 & 5.378e+00 & 3.850e+00 & 4.327e+01 & 4.482e+01 \\
6.086e+02 & 1.437e+00 & 2.620e+00 & 5.478e+00 & 3.822e+00 & 4.281e+01 & 4.447e+01 \\
6.110e+02 & 1.440e+00 & 2.745e+00 & 5.578e+00 & 3.784e+00 & 4.209e+01 & 4.389e+01 \\
6.577e+02 & 1.508e+00 & 2.941e+00 & 5.679e+00 & 3.711e+00 & 4.035e+01 & 4.259e+01 \\
7.210e+02 & 1.586e+00 & 3.007e+00 & 5.790e+00 & 3.706e+00 & 4.035e+01 & 4.255e+01 \\
8.122e+02 & 1.678e+00 & 3.057e+00 & 5.891e+00 & 3.706e+00 & 4.045e+01 & 4.265e+01 \\
9.100e+02 & 1.759e+00 & 3.096e+00 & 5.993e+00 & 3.712e+00 & 4.078e+01 & 4.290e+01 \\
1.036e+03 & 1.846e+00 & 3.153e+00 & 6.093e+00 & 3.709e+00 & 4.077e+01 & 4.293e+01 \\
1.183e+03 & 1.928e+00 & 3.198e+00 & 6.193e+00 & 3.711e+00 & 4.087e+01 & 4.311e+01 \\
1.356e+03 & 2.009e+00 & 3.241e+00 & 6.293e+00 & 3.715e+00 & 4.108e+01 & 4.328e+01 \\
1.554e+03 & 2.086e+00 & 3.283e+00 & 6.394e+00 & 3.719e+00 & 4.129e+01 & 4.345e+01 \\
1.776e+03 & 2.159e+00 & 3.326e+00 & 6.494e+00 & 3.723e+00 & 4.151e+01 & 4.363e+01 \\
2.037e+03 & 2.231e+00 & 3.369e+00 & 6.594e+00 & 3.726e+00 & 4.171e+01 & 4.380e+01 \\
2.354e+03 & 2.305e+00 & 3.415e+00 & 6.694e+00 & 3.728e+00 & 4.181e+01 & 4.397e+01 \\
2.753e+03 & 2.383e+00 & 3.460e+00 & 6.794e+00 & 3.731e+00 & 4.202e+01 & 4.407e+01 \\
3.258e+03 & 2.466e+00 & 3.504e+00 & 6.894e+00 & 3.734e+00 & 4.222e+01 & 4.423e+01 \\
3.886e+03 & 2.550e+00 & 3.549e+00 & 6.994e+00 & 3.736e+00 & 4.232e+01 & 4.440e+01 \\
4.658e+03 & 2.636e+00 & 3.595e+00 & 7.094e+00 & 3.738e+00 & 4.253e+01 & 4.457e+01 \\
5.624e+03 & 2.723e+00 & 3.640e+00 & 7.194e+00 & 3.740e+00 & 4.273e+01 & 4.467e+01 \\
6.828e+03 & 2.812e+00 & 3.685e+00 & 7.294e+00 & 3.743e+00 & 4.282e+01 & 4.484e+01 \\
8.331e+03 & 2.903e+00 & 3.729e+00 & 7.395e+00 & 3.746e+00 & 4.302e+01 & 4.500e+01 \\
1.021e+04 & 2.995e+00 & 3.775e+00 & 7.495e+00 & 3.748e+00 & 4.312e+01 & 4.517e+01 \\
1.235e+04 & 3.080e+00 & 3.826e+00 & 7.601e+00 & 3.749e+00 & 4.333e+01 & 4.527e+01 \\
1.568e+04 & 3.186e+00 & 3.863e+00 & 7.701e+00 & 3.756e+00 & 4.352e+01 & 4.550e+01 \\
1.940e+04 & 3.280e+00 & 3.911e+00 & 7.802e+00 & 3.757e+00 & 4.371e+01 & 4.559e+01 \\
2.406e+04 & 3.375e+00 & 3.956e+00 & 7.902e+00 & 3.760e+00 & 4.381e+01 & 4.576e+01 \\
2.986e+04 & 3.470e+00 & 3.998e+00 & 8.002e+00 & 3.763e+00 & 4.400e+01 & 4.598e+01 \\
3.716e+04 & 3.566e+00 & 4.041e+00 & 8.102e+00 & 3.767e+00 & 4.419e+01 & 4.613e+01 \\
4.632e+04 & 3.663e+00 & 4.084e+00 & 8.202e+00 & 3.771e+00 & 4.438e+01 & 4.629e+01 \\
5.778e+04 & 3.759e+00 & 4.125e+00 & 8.302e+00 & 3.775e+00 & 4.465e+01 & 4.644e+01 \\
7.218e+04 & 3.856e+00 & 4.167e+00 & 8.402e+00 & 3.779e+00 & 4.484e+01 & 4.666e+01 \\
9.023e+04 & 3.954e+00 & 4.207e+00 & 8.502e+00 & 3.784e+00 & 4.502e+01 & 4.681e+01 \\
1.129e+05 & 4.051e+00 & 4.247e+00 & 8.602e+00 & 3.789e+00 & 4.520e+01 & 4.702e+01 \\
1.413e+05 & 4.149e+00 & 4.287e+00 & 8.702e+00 & 3.794e+00 & 4.545e+01 & 4.722e+01 \\
1.770e+05 & 4.247e+00 & 4.325e+00 & 8.802e+00 & 3.800e+00 & 4.571e+01 & 4.742e+01 \\
2.219e+05 & 4.345e+00 & 4.362e+00 & 8.902e+00 & 3.807e+00 & 4.595e+01 & 4.761e+01 \\
2.782e+05 & 4.444e+00 & 4.398e+00 & 9.002e+00 & 3.814e+00 & 4.619e+01 & 4.785e+01 \\
3.490e+05 & 4.542e+00 & 4.429e+00 & 9.102e+00 & 3.823e+00 & 4.650e+01 & 4.809e+01 \\
4.381e+05 & 4.641e+00 & 4.453e+00 & 9.202e+00 & 3.836e+00 & 4.686e+01 & 4.840e+01 \\
5.498e+05 & 4.740e+00 & 4.469e+00 & 9.302e+00 & 3.853e+00 & 4.725e+01 & 4.878e+01 \\
6.713e+05 & 4.827e+00 & 4.462e+00 & 9.388e+00 & 3.878e+00 & 4.783e+01 & 4.921e+01 \\
\hline\end{tabular}\end{center}\label{tab-1em1}\end{table*}

\begin{table*}\caption{Model at $\dm=0.01$ \Mpy}\begin{center}\begin{tabular}{|c|c|c|c|c|c|c|}\hline
age [yr] & log($M$/\Ms) & log($R$/\Rs) & log($L$/\Ls) & log(\Teff[K]) & log(\sion[s$^{-1}$]) & log(\slw[s$^{-1}$]) \\\hline
2.100e+02 & 3.222e-01 & 1.381e+00 & 2.499e+00 & 3.696e+00 & 3.682e+01 & 3.901e+01 \\
4.030e+02 & 6.053e-01 & 1.285e+00 & 2.408e+00 & 3.722e+00 & 3.742e+01 & 3.954e+01 \\
7.284e+02 & 8.623e-01 & 1.193e+00 & 2.325e+00 & 3.747e+00 & 3.795e+01 & 3.993e+01 \\
1.173e+03 & 1.069e+00 & 1.197e+00 & 2.425e+00 & 3.770e+00 & 3.860e+01 & 4.052e+01 \\
1.192e+03 & 1.076e+00 & 1.240e+00 & 2.562e+00 & 3.782e+00 & 3.908e+01 & 4.087e+01 \\
1.192e+03 & 1.076e+00 & 1.248e+00 & 2.679e+00 & 3.808e+00 & 3.973e+01 & 4.144e+01 \\
1.192e+03 & 1.076e+00 & 1.256e+00 & 2.788e+00 & 3.831e+00 & 4.031e+01 & 4.190e+01 \\
1.193e+03 & 1.077e+00 & 1.260e+00 & 2.893e+00 & 3.855e+00 & 4.090e+01 & 4.240e+01 \\
1.193e+03 & 1.077e+00 & 1.261e+00 & 2.999e+00 & 3.881e+00 & 4.149e+01 & 4.289e+01 \\
1.193e+03 & 1.077e+00 & 1.248e+00 & 3.085e+00 & 3.909e+00 & 4.204e+01 & 4.334e+01 \\
1.193e+03 & 1.077e+00 & 1.273e+00 & 3.195e+00 & 3.924e+00 & 4.240e+01 & 4.364e+01 \\
1.194e+03 & 1.077e+00 & 1.277e+00 & 3.307e+00 & 3.950e+00 & 4.289e+01 & 4.404e+01 \\
1.194e+03 & 1.077e+00 & 1.281e+00 & 3.407e+00 & 3.973e+00 & 4.333e+01 & 4.440e+01 \\
1.194e+03 & 1.077e+00 & 1.288e+00 & 3.508e+00 & 3.995e+00 & 4.370e+01 & 4.473e+01 \\
1.195e+03 & 1.077e+00 & 1.292e+00 & 3.615e+00 & 4.020e+00 & 4.413e+01 & 4.507e+01 \\
1.195e+03 & 1.077e+00 & 1.296e+00 & 3.718e+00 & 4.043e+00 & 4.451e+01 & 4.538e+01 \\
1.196e+03 & 1.078e+00 & 1.301e+00 & 3.819e+00 & 4.066e+00 & 4.484e+01 & 4.568e+01 \\
1.196e+03 & 1.078e+00 & 1.310e+00 & 3.938e+00 & 4.091e+00 & 4.522e+01 & 4.599e+01 \\
1.197e+03 & 1.078e+00 & 1.319e+00 & 4.041e+00 & 4.113e+00 & 4.554e+01 & 4.624e+01 \\
1.198e+03 & 1.079e+00 & 1.330e+00 & 4.144e+00 & 4.133e+00 & 4.582e+01 & 4.647e+01 \\
1.200e+03 & 1.079e+00 & 1.346e+00 & 4.248e+00 & 4.151e+00 & 4.608e+01 & 4.668e+01 \\
1.202e+03 & 1.080e+00 & 1.362e+00 & 4.352e+00 & 4.169e+00 & 4.633e+01 & 4.689e+01 \\
1.206e+03 & 1.081e+00 & 1.384e+00 & 4.453e+00 & 4.183e+00 & 4.653e+01 & 4.706e+01 \\
1.212e+03 & 1.084e+00 & 1.412e+00 & 4.559e+00 & 4.196e+00 & 4.673e+01 & 4.723e+01 \\
1.223e+03 & 1.088e+00 & 1.452e+00 & 4.659e+00 & 4.201e+00 & 4.687e+01 & 4.736e+01 \\
1.240e+03 & 1.094e+00 & 1.497e+00 & 4.761e+00 & 4.204e+00 & 4.699e+01 & 4.747e+01 \\
1.269e+03 & 1.103e+00 & 1.561e+00 & 4.861e+00 & 4.197e+00 & 4.704e+01 & 4.754e+01 \\
1.316e+03 & 1.119e+00 & 1.643e+00 & 4.962e+00 & 4.181e+00 & 4.703e+01 & 4.756e+01 \\
1.392e+03 & 1.144e+00 & 1.743e+00 & 5.062e+00 & 4.156e+00 & 4.693e+01 & 4.753e+01 \\
1.512e+03 & 1.180e+00 & 1.850e+00 & 5.162e+00 & 4.128e+00 & 4.679e+01 & 4.745e+01 \\
1.700e+03 & 1.231e+00 & 1.933e+00 & 5.262e+00 & 4.111e+00 & 4.674e+01 & 4.745e+01 \\
1.965e+03 & 1.293e+00 & 1.963e+00 & 5.362e+00 & 4.121e+00 & 4.693e+01 & 4.761e+01 \\
2.210e+03 & 1.344e+00 & 1.950e+00 & 5.436e+00 & 4.146e+00 & 4.722e+01 & 4.784e+01 \\
2.419e+03 & 1.384e+00 & 1.928e+00 & 5.493e+00 & 4.171e+00 & 4.748e+01 & 4.804e+01 \\
2.624e+03 & 1.419e+00 & 1.904e+00 & 5.544e+00 & 4.196e+00 & 4.772e+01 & 4.822e+01 \\
2.833e+03 & 1.452e+00 & 1.878e+00 & 5.593e+00 & 4.221e+00 & 4.794e+01 & 4.839e+01 \\
3.054e+03 & 1.485e+00 & 1.852e+00 & 5.642e+00 & 4.246e+00 & 4.815e+01 & 4.854e+01 \\
3.292e+03 & 1.517e+00 & 1.827e+00 & 5.691e+00 & 4.271e+00 & 4.835e+01 & 4.869e+01 \\
3.548e+03 & 1.550e+00 & 1.802e+00 & 5.741e+00 & 4.296e+00 & 4.853e+01 & 4.882e+01 \\
3.833e+03 & 1.584e+00 & 1.778e+00 & 5.793e+00 & 4.321e+00 & 4.871e+01 & 4.894e+01 \\
4.152e+03 & 1.618e+00 & 1.755e+00 & 5.848e+00 & 4.346e+00 & 4.888e+01 & 4.906e+01 \\
4.525e+03 & 1.656e+00 & 1.735e+00 & 5.906e+00 & 4.371e+00 & 4.904e+01 & 4.917e+01 \\
4.980e+03 & 1.697e+00 & 1.718e+00 & 5.972e+00 & 4.396e+00 & 4.920e+01 & 4.929e+01 \\
5.609e+03 & 1.749e+00 & 1.708e+00 & 6.054e+00 & 4.421e+00 & 4.937e+01 & 4.941e+01 \\
6.495e+03 & 1.813e+00 & 1.722e+00 & 6.154e+00 & 4.439e+00 & 4.953e+01 & 4.954e+01 \\
7.541e+03 & 1.877e+00 & 1.779e+00 & 6.254e+00 & 4.436e+00 & 4.962e+01 & 4.963e+01 \\
8.809e+03 & 1.945e+00 & 1.905e+00 & 6.355e+00 & 4.398e+00 & 4.959e+01 & 4.967e+01 \\
1.048e+04 & 2.020e+00 & 2.105e+00 & 6.455e+00 & 4.323e+00 & 4.938e+01 & 4.961e+01 \\
1.279e+04 & 2.107e+00 & 2.214e+00 & 6.555e+00 & 4.293e+00 & 4.933e+01 & 4.962e+01 \\
1.392e+04 & 2.144e+00 & 2.184e+00 & 6.594e+00 & 4.318e+00 & 4.950e+01 & 4.974e+01 \\
1.470e+04 & 2.167e+00 & 2.146e+00 & 6.618e+00 & 4.344e+00 & 4.964e+01 & 4.982e+01 \\
1.534e+04 & 2.186e+00 & 2.105e+00 & 6.637e+00 & 4.369e+00 & 4.976e+01 & 4.990e+01 \\
1.596e+04 & 2.203e+00 & 2.064e+00 & 6.655e+00 & 4.394e+00 & 4.988e+01 & 4.997e+01 \\
1.659e+04 & 2.220e+00 & 2.022e+00 & 6.671e+00 & 4.419e+00 & 4.998e+01 & 5.002e+01 \\
1.735e+04 & 2.239e+00 & 1.983e+00 & 6.694e+00 & 4.444e+00 & 5.008e+01 & 5.008e+01 \\
1.812e+04 & 2.258e+00 & 1.943e+00 & 6.714e+00 & 4.469e+00 & 5.017e+01 & 5.013e+01 \\
1.895e+04 & 2.278e+00 & 1.903e+00 & 6.735e+00 & 4.494e+00 & 5.026e+01 & 5.017e+01 \\
2.013e+04 & 2.304e+00 & 1.868e+00 & 6.765e+00 & 4.519e+00 & 5.034e+01 & 5.021e+01 \\
2.182e+04 & 2.339e+00 & 1.839e+00 & 6.806e+00 & 4.544e+00 & 5.044e+01 & 5.025e+01 \\
2.619e+04 & 2.418e+00 & 1.949e+00 & 6.906e+00 & 4.514e+00 & 5.048e+01 & 5.034e+01 \\
3.151e+04 & 2.499e+00 & 2.622e+00 & 7.006e+00 & 4.203e+00 & 4.922e+01 & 4.971e+01 \\
3.527e+04 & 2.547e+00 & 2.596e+00 & 7.055e+00 & 4.228e+00 & 4.944e+01 & 4.988e+01 \\
3.583e+04 & 2.554e+00 & 2.549e+00 & 7.061e+00 & 4.253e+00 & 4.961e+01 & 4.999e+01
\end{tabular}\end{center}\label{tab-1em2}\end{table*}
\begin{table*}\contcaption{}\begin{center}\begin{tabular}{|c|c|c|c|c|c|c|}
3.639e+04 & 2.561e+00 & 2.502e+00 & 7.068e+00 & 4.278e+00 & 4.976e+01 & 5.008e+01 \\
3.694e+04 & 2.567e+00 & 2.455e+00 & 7.074e+00 & 4.303e+00 & 4.990e+01 & 5.017e+01 \\
3.753e+04 & 2.574e+00 & 2.408e+00 & 7.080e+00 & 4.328e+00 & 5.003e+01 & 5.025e+01 \\
3.818e+04 & 2.582e+00 & 2.362e+00 & 7.087e+00 & 4.353e+00 & 5.014e+01 & 5.032e+01 \\
3.884e+04 & 2.589e+00 & 2.315e+00 & 7.095e+00 & 4.378e+00 & 5.025e+01 & 5.038e+01 \\
3.958e+04 & 2.597e+00 & 2.269e+00 & 7.103e+00 & 4.403e+00 & 5.035e+01 & 5.043e+01 \\
4.038e+04 & 2.606e+00 & 2.224e+00 & 7.111e+00 & 4.428e+00 & 5.045e+01 & 5.048e+01 \\
4.127e+04 & 2.616e+00 & 2.178e+00 & 7.121e+00 & 4.453e+00 & 5.053e+01 & 5.052e+01 \\
4.229e+04 & 2.626e+00 & 2.134e+00 & 7.131e+00 & 4.478e+00 & 5.061e+01 & 5.055e+01 \\
4.341e+04 & 2.638e+00 & 2.089e+00 & 7.143e+00 & 4.503e+00 & 5.069e+01 & 5.058e+01 \\
4.480e+04 & 2.651e+00 & 2.046e+00 & 7.157e+00 & 4.528e+00 & 5.076e+01 & 5.060e+01 \\
4.665e+04 & 2.669e+00 & 2.006e+00 & 7.176e+00 & 4.553e+00 & 5.082e+01 & 5.062e+01 \\
5.608e+04 & 2.749e+00 & 2.406e+00 & 7.276e+00 & 4.378e+00 & 5.044e+01 & 5.056e+01 \\
6.911e+04 & 2.840e+00 & 3.702e+00 & 7.376e+00 & 3.755e+00 & 4.319e+01 & 4.517e+01 \\
8.395e+04 & 2.924e+00 & 3.753e+00 & 7.476e+00 & 3.754e+00 & 4.330e+01 & 4.527e+01 \\
1.056e+05 & 3.024e+00 & 3.778e+00 & 7.579e+00 & 3.768e+00 & 4.376e+01 & 4.561e+01 \\
1.381e+05 & 3.140e+00 & 3.848e+00 & 7.679e+00 & 3.758e+00 & 4.359e+01 & 4.553e+01 \\
1.458e+05 & 3.164e+00 & 3.803e+00 & 7.699e+00 & 3.785e+00 & 4.421e+01 & 4.606e+01 \\
1.695e+05 & 3.229e+00 & 3.907e+00 & 7.799e+00 & 3.758e+00 & 4.371e+01 & 4.565e+01 \\
2.232e+05 & 3.349e+00 & 3.948e+00 & 7.899e+00 & 3.763e+00 & 4.390e+01 & 4.581e+01 \\
2.777e+05 & 3.444e+00 & 3.987e+00 & 7.999e+00 & 3.768e+00 & 4.418e+01 & 4.603e+01 \\
3.505e+05 & 3.545e+00 & 4.032e+00 & 8.099e+00 & 3.771e+00 & 4.428e+01 & 4.619e+01 \\
4.413e+05 & 3.645e+00 & 4.071e+00 & 8.199e+00 & 3.776e+00 & 4.455e+01 & 4.640e+01 \\
5.442e+05 & 3.736e+00 & 4.087e+00 & 8.299e+00 & 3.793e+00 & 4.505e+01 & 4.682e+01 \\
6.905e+05 & 3.839e+00 & 4.149e+00 & 8.399e+00 & 3.787e+00 & 4.499e+01 & 4.682e+01 \\
8.634e+05 & 3.936e+00 & 4.185e+00 & 8.499e+00 & 3.794e+00 & 4.525e+01 & 4.702e+01 \\
1.084e+06 & 4.035e+00 & 4.223e+00 & 8.599e+00 & 3.800e+00 & 4.550e+01 & 4.721e+01 \\
1.361e+06 & 4.134e+00 & 4.255e+00 & 8.699e+00 & 3.809e+00 & 4.575e+01 & 4.746e+01 \\
\hline\end{tabular}\end{center}\label{tab-1em2}\end{table*}

\begin{table*}\caption{Model at $\dm=0.001$ \Mpy}\begin{center}\begin{tabular}{|c|c|c|c|c|c|c|}\hline
age [yr] & log($M$/\Ms) & log($R$/\Rs) & log($L$/\Ls) & log(\Teff[K]) & log(\sion[s$^{-1}$]) & log(\slw[s$^{-1}$]) \\\hline
2.010e+03 & 3.032e-01 & 1.368e+00 & 2.474e+00 & 3.696e+00 & 3.679e+01 & 3.899e+01 \\
3.672e+03 & 5.649e-01 & 1.239e+00 & 2.316e+00 & 3.721e+00 & 3.733e+01 & 3.945e+01 \\
6.607e+03 & 8.200e-01 & 1.186e+00 & 2.309e+00 & 3.746e+00 & 3.794e+01 & 3.991e+01 \\
7.180e+03 & 8.561e-01 & 1.227e+00 & 2.410e+00 & 3.751e+00 & 3.813e+01 & 4.014e+01 \\
7.323e+03 & 8.647e-01 & 1.270e+00 & 2.513e+00 & 3.755e+00 & 3.833e+01 & 4.031e+01 \\
7.340e+03 & 8.657e-01 & 1.311e+00 & 2.635e+00 & 3.765e+00 & 3.873e+01 & 4.061e+01 \\
7.348e+03 & 8.662e-01 & 1.343e+00 & 2.752e+00 & 3.778e+00 & 3.919e+01 & 4.101e+01 \\
7.357e+03 & 8.667e-01 & 1.356e+00 & 2.853e+00 & 3.797e+00 & 3.968e+01 & 4.142e+01 \\
7.369e+03 & 8.674e-01 & 1.368e+00 & 2.966e+00 & 3.820e+00 & 4.030e+01 & 4.191e+01 \\
7.384e+03 & 8.683e-01 & 1.380e+00 & 3.073e+00 & 3.840e+00 & 4.079e+01 & 4.235e+01 \\
7.405e+03 & 8.695e-01 & 1.393e+00 & 3.181e+00 & 3.861e+00 & 4.130e+01 & 4.276e+01 \\
7.436e+03 & 8.713e-01 & 1.408e+00 & 3.290e+00 & 3.881e+00 & 4.173e+01 & 4.315e+01 \\
7.479e+03 & 8.738e-01 & 1.424e+00 & 3.393e+00 & 3.899e+00 & 4.217e+01 & 4.350e+01 \\
7.546e+03 & 8.777e-01 & 1.442e+00 & 3.496e+00 & 3.915e+00 & 4.254e+01 & 4.384e+01 \\
7.657e+03 & 8.841e-01 & 1.462e+00 & 3.602e+00 & 3.932e+00 & 4.289e+01 & 4.412e+01 \\
7.828e+03 & 8.937e-01 & 1.481e+00 & 3.703e+00 & 3.947e+00 & 4.325e+01 & 4.442e+01 \\
8.097e+03 & 9.083e-01 & 1.497e+00 & 3.805e+00 & 3.965e+00 & 4.360e+01 & 4.472e+01 \\
8.493e+03 & 9.291e-01 & 1.503e+00 & 3.905e+00 & 3.987e+00 & 4.398e+01 & 4.503e+01 \\
8.972e+03 & 9.529e-01 & 1.498e+00 & 3.995e+00 & 4.012e+00 & 4.440e+01 & 4.539e+01 \\
9.458e+03 & 9.758e-01 & 1.484e+00 & 4.068e+00 & 4.037e+00 & 4.477e+01 & 4.569e+01 \\
9.939e+03 & 9.973e-01 & 1.466e+00 & 4.133e+00 & 4.062e+00 & 4.512e+01 & 4.595e+01 \\
1.043e+04 & 1.018e+00 & 1.446e+00 & 4.192e+00 & 4.087e+00 & 4.544e+01 & 4.621e+01 \\
1.092e+04 & 1.038e+00 & 1.424e+00 & 4.248e+00 & 4.112e+00 & 4.573e+01 & 4.645e+01 \\
1.143e+04 & 1.058e+00 & 1.400e+00 & 4.302e+00 & 4.137e+00 & 4.602e+01 & 4.665e+01 \\
1.194e+04 & 1.077e+00 & 1.376e+00 & 4.355e+00 & 4.163e+00 & 4.627e+01 & 4.685e+01 \\
1.247e+04 & 1.096e+00 & 1.351e+00 & 4.406e+00 & 4.188e+00 & 4.652e+01 & 4.705e+01 \\
1.302e+04 & 1.114e+00 & 1.325e+00 & 4.455e+00 & 4.213e+00 & 4.675e+01 & 4.721e+01 \\
1.357e+04 & 1.132e+00 & 1.299e+00 & 4.503e+00 & 4.238e+00 & 4.697e+01 & 4.737e+01 \\
1.414e+04 & 1.150e+00 & 1.272e+00 & 4.551e+00 & 4.264e+00 & 4.717e+01 & 4.752e+01 \\
1.473e+04 & 1.168e+00 & 1.244e+00 & 4.597e+00 & 4.289e+00 & 4.735e+01 & 4.765e+01 \\
1.534e+04 & 1.186e+00 & 1.217e+00 & 4.642e+00 & 4.314e+00 & 4.752e+01 & 4.777e+01 \\
1.596e+04 & 1.203e+00 & 1.189e+00 & 4.686e+00 & 4.339e+00 & 4.768e+01 & 4.789e+01 \\
1.662e+04 & 1.221e+00 & 1.160e+00 & 4.730e+00 & 4.365e+00 & 4.783e+01 & 4.799e+01 \\
1.731e+04 & 1.238e+00 & 1.131e+00 & 4.774e+00 & 4.390e+00 & 4.798e+01 & 4.808e+01 \\
1.803e+04 & 1.256e+00 & 1.102e+00 & 4.818e+00 & 4.416e+00 & 4.811e+01 & 4.817e+01 \\
1.878e+04 & 1.274e+00 & 1.073e+00 & 4.861e+00 & 4.441e+00 & 4.824e+01 & 4.824e+01 \\
1.956e+04 & 1.291e+00 & 1.043e+00 & 4.904e+00 & 4.466e+00 & 4.836e+01 & 4.831e+01 \\
2.036e+04 & 1.309e+00 & 1.014e+00 & 4.947e+00 & 4.492e+00 & 4.846e+01 & 4.837e+01 \\
2.121e+04 & 1.327e+00 & 9.848e-01 & 4.989e+00 & 4.517e+00 & 4.856e+01 & 4.843e+01 \\
2.212e+04 & 1.345e+00 & 9.548e-01 & 5.030e+00 & 4.542e+00 & 4.866e+01 & 4.847e+01 \\
2.307e+04 & 1.363e+00 & 9.244e-01 & 5.072e+00 & 4.568e+00 & 4.875e+01 & 4.852e+01 \\
2.407e+04 & 1.381e+00 & 8.949e-01 & 5.113e+00 & 4.593e+00 & 4.883e+01 & 4.856e+01 \\
2.512e+04 & 1.400e+00 & 8.653e-01 & 5.154e+00 & 4.618e+00 & 4.890e+01 & 4.859e+01 \\
2.622e+04 & 1.419e+00 & 8.358e-01 & 5.196e+00 & 4.643e+00 & 4.898e+01 & 4.862e+01 \\
2.738e+04 & 1.437e+00 & 8.068e-01 & 5.239e+00 & 4.668e+00 & 4.905e+01 & 4.864e+01 \\
2.860e+04 & 1.456e+00 & 7.781e-01 & 5.282e+00 & 4.694e+00 & 4.911e+01 & 4.867e+01 \\
2.988e+04 & 1.475e+00 & 7.504e-01 & 5.328e+00 & 4.719e+00 & 4.918e+01 & 4.869e+01 \\
3.120e+04 & 1.494e+00 & 7.232e-01 & 5.375e+00 & 4.744e+00 & 4.924e+01 & 4.870e+01 \\
3.258e+04 & 1.513e+00 & 6.960e-01 & 5.420e+00 & 4.769e+00 & 4.929e+01 & 4.872e+01 \\
3.403e+04 & 1.532e+00 & 6.686e-01 & 5.466e+00 & 4.794e+00 & 4.935e+01 & 4.873e+01 \\
3.557e+04 & 1.551e+00 & 6.411e-01 & 5.511e+00 & 4.819e+00 & 4.940e+01 & 4.874e+01 \\
3.721e+04 & 1.571e+00 & 6.138e-01 & 5.557e+00 & 4.844e+00 & 4.945e+01 & 4.874e+01 \\
3.899e+04 & 1.591e+00 & 5.866e-01 & 5.603e+00 & 4.870e+00 & 4.949e+01 & 4.874e+01 \\
4.094e+04 & 1.612e+00 & 5.591e-01 & 5.649e+00 & 4.895e+00 & 4.954e+01 & 4.875e+01 \\
4.543e+04 & 1.657e+00 & 5.374e-01 & 5.705e+00 & 4.920e+00 & 4.959e+01 & 4.876e+01 \\
5.472e+04 & 1.738e+00 & 5.057e-01 & 5.742e+00 & 4.945e+00 & 4.962e+01 & 4.874e+01 \\
6.636e+04 & 1.822e+00 & 5.357e-01 & 5.842e+00 & 4.955e+00 & 4.972e+01 & 4.883e+01 \\
7.631e+04 & 1.883e+00 & 5.691e-01 & 5.942e+00 & 4.963e+00 & 4.982e+01 & 4.891e+01 \\
8.784e+04 & 1.944e+00 & 6.047e-01 & 6.042e+00 & 4.970e+00 & 4.991e+01 & 4.899e+01 \\
1.014e+05 & 2.006e+00 & 6.414e-01 & 6.142e+00 & 4.977e+00 & 5.001e+01 & 4.908e+01 \\
1.173e+05 & 2.069e+00 & 6.798e-01 & 6.243e+00 & 4.983e+00 & 5.011e+01 & 4.917e+01 \\
1.363e+05 & 2.135e+00 & 7.194e-01 & 6.343e+00 & 4.988e+00 & 5.021e+01 & 4.926e+01 \\
1.591e+05 & 2.202e+00 & 7.600e-01 & 6.443e+00 & 4.993e+00 & 5.031e+01 & 4.935e+01
\end{tabular}\end{center}\label{tab-1em3}\end{table*}
\begin{table*}\contcaption{}\begin{center}\begin{tabular}{|c|c|c|c|c|c|c|}
1.864e+05 & 2.270e+00 & 8.003e-01 & 6.543e+00 & 4.998e+00 & 5.041e+01 & 4.944e+01 \\
2.194e+05 & 2.341e+00 & 8.408e-01 & 6.643e+00 & 5.002e+00 & 5.050e+01 & 4.953e+01 \\
2.596e+05 & 2.414e+00 & 8.827e-01 & 6.743e+00 & 5.006e+00 & 5.060e+01 & 4.962e+01 \\
3.083e+05 & 2.489e+00 & 9.281e-01 & 6.843e+00 & 5.009e+00 & 5.070e+01 & 4.971e+01 \\
3.678e+05 & 2.566e+00 & 9.758e-01 & 6.943e+00 & 5.010e+00 & 5.080e+01 & 4.981e+01 \\
4.403e+05 & 2.644e+00 & 1.023e+00 & 7.043e+00 & 5.011e+00 & 5.090e+01 & 4.991e+01 \\
5.289e+05 & 2.723e+00 & 1.075e+00 & 7.143e+00 & 5.010e+00 & 5.100e+01 & 5.001e+01 \\
6.373e+05 & 2.804e+00 & 1.128e+00 & 7.243e+00 & 5.009e+00 & 5.110e+01 & 5.011e+01 \\
7.688e+05 & 2.886e+00 & 1.186e+00 & 7.343e+00 & 5.005e+00 & 5.120e+01 & 5.022e+01 \\
9.324e+05 & 2.970e+00 & 1.247e+00 & 7.443e+00 & 4.999e+00 & 5.131e+01 & 5.033e+01 \\
1.137e+06 & 3.056e+00 & 1.309e+00 & 7.543e+00 & 4.993e+00 & 5.141e+01 & 5.045e+01 \\
1.388e+06 & 3.142e+00 & 1.382e+00 & 7.643e+00 & 4.982e+00 & 5.151e+01 & 5.057e+01 \\
1.714e+06 & 3.234e+00 & 1.446e+00 & 7.743e+00 & 4.975e+00 & 5.161e+01 & 5.069e+01 \\
2.111e+06 & 3.325e+00 & 1.513e+00 & 7.844e+00 & 4.966e+00 & 5.172e+01 & 5.080e+01 \\
2.609e+06 & 3.416e+00 & 1.585e+00 & 7.944e+00 & 4.955e+00 & 5.182e+01 & 5.092e+01 \\
3.228e+06 & 3.509e+00 & 1.676e+00 & 8.044e+00 & 4.935e+00 & 5.193e+01 & 5.107e+01 \\
4.011e+06 & 3.603e+00 & 1.769e+00 & 8.144e+00 & 4.913e+00 & 5.203e+01 & 5.121e+01 \\
4.989e+06 & 3.698e+00 & 1.870e+00 & 8.244e+00 & 4.888e+00 & 5.213e+01 & 5.135e+01 \\
6.220e+06 & 3.794e+00 & 1.936e+00 & 8.344e+00 & 4.880e+00 & 5.223e+01 & 5.147e+01 \\
7.756e+06 & 3.890e+00 & 2.092e+00 & 8.444e+00 & 4.827e+00 & 5.233e+01 & 5.166e+01 \\
8.579e+06 & 3.933e+00 & 2.065e+00 & 8.489e+00 & 4.852e+00 & 5.238e+01 & 5.166e+01 \\
1.073e+07 & 4.030e+00 & 2.281e+00 & 8.589e+00 & 4.769e+00 & 5.246e+01 & 5.189e+01 \\
1.164e+07 & 4.066e+00 & 2.278e+00 & 8.684e+00 & 4.794e+00 & 5.257e+01 & 5.195e+01 \\
\hline\end{tabular}\end{center}\label{tab-1em3}\end{table*}

\section*{Acknowledgements}

LH, RSK and DJW were supported by the European Research Council under the European Community's Seventh Framework Programme (FP7/2007 - 2013)
via the ERC Advanced Grant `STARLIGHT: Formation of the First Stars' (project number 339177).
Part of this work was supported by the Swiss National Science Foundation.
DJW was supported by STFC New Applicant Grant ST/P000509/1.


\bibliographystyle{mnras}
\bibliography{bibliotheque}

\begin{thebibliography}{}
\makeatletter
\relax
\def\mn@urlcharsother{\let\do\@makeother \do\$\do\&\do\#\do\^\do\_\do\%\do\~}
\def\mn@doi{\begingroup\mn@urlcharsother \@ifnextchar [ {\mn@doi@}
  {\mn@doi@[]}}
\def\mn@doi@[#1]#2{\def\@tempa{#1}\ifx\@tempa\@empty \href
  {http://dx.doi.org/#2} {doi:#2}\else \href {http://dx.doi.org/#2} {#1}\fi
  \endgroup}
\def\mn@eprint#1#2{\mn@eprint@#1:#2::\@nil}
\def\mn@eprint@arXiv#1{\href {http://arxiv.org/abs/#1} {{\tt arXiv:#1}}}
\def\mn@eprint@dblp#1{\href {http://dblp.uni-trier.de/rec/bibtex/#1.xml}
  {dblp:#1}}
\def\mn@eprint@#1:#2:#3:#4\@nil{\def\@tempa {#1}\def\@tempb {#2}\def\@tempc
  {#3}\ifx \@tempc \@empty \let \@tempc \@tempb \let \@tempb \@tempa \fi \ifx
  \@tempb \@empty \def\@tempb {arXiv}\fi \@ifundefined
  {mn@eprint@\@tempb}{\@tempb:\@tempc}{\expandafter \expandafter \csname
  mn@eprint@\@tempb\endcsname \expandafter{\@tempc}}}

\bibitem[\protect\citeauthoryear{{Agarwal}, {Khochfar}, {Johnson}, {Neistein},
  {Dalla Vecchia}  \& {Livio}}{{Agarwal} et~al.}{2012}]{agarwal2012}
{Agarwal} B.,  {Khochfar} S.,  {Johnson} J.~L.,  {Neistein} E.,  {Dalla
  Vecchia} C.,   {Livio} M.,  2012, \mn@doi [\mnras]
  {10.1111/j.1365-2966.2012.21651.x}, \href
  {http://cdsads.u-strasbg.fr/abs/2012MNRAS.425.2854A} {425, 2854}

\bibitem[\protect\citeauthoryear{{Agarwal}, {Johnson}, {Zackrisson}, {Labbe},
  {van den Bosch}, {Natarajan}  \& {Khochfar}}{{Agarwal}
  et~al.}{2016}]{agarwal2016b}
{Agarwal} B.,  {Johnson} J.~L.,  {Zackrisson} E.,  {Labbe} I.,  {van den Bosch}
  F.~C.,  {Natarajan} P.,   {Khochfar} S.,  2016, \mn@doi [\mnras]
  {10.1093/mnras/stw1173}, \href
  {http://cdsads.u-strasbg.fr/abs/2016MNRAS.460.4003A} {460, 4003}

\bibitem[\protect\citeauthoryear{{Alvarez}, {Wise}  \& {Abel}}{{Alvarez}
  et~al.}{2009}]{alvarez2009}
{Alvarez} M.~A.,  {Wise} J.~H.,   {Abel} T.,  2009, \mn@doi [\apjl]
  {10.1088/0004-637X/701/2/L133}, \href
  {http://cdsads.u-strasbg.fr/abs/2009ApJ...701L.133A} {701, L133}

\bibitem[\protect\citeauthoryear{{Appenzeller} \& {Fricke}}{{Appenzeller} \&
  {Fricke}}{1972a}]{appenzeller1972a}
{Appenzeller} I.,  {Fricke} K.,  1972a, \aap, \href
  {http://cdsads.u-strasbg.fr/abs/1972A%26A....18...10A} {18, 10}

\bibitem[\protect\citeauthoryear{{Appenzeller} \& {Fricke}}{{Appenzeller} \&
  {Fricke}}{1972b}]{appenzeller1972b}
{Appenzeller} I.,  {Fricke} K.,  1972b, \aap, \href
  {http://cdsads.u-strasbg.fr/abs/1972A%26A....21..285A} {21, 285}

\bibitem[\protect\citeauthoryear{{Becerra}, {Greif}, {Springel}  \&
  {Hernquist}}{{Becerra} et~al.}{2015}]{becerra2015}
{Becerra} F.,  {Greif} T.~H.,  {Springel} V.,   {Hernquist} L.~E.,  2015,
  \mn@doi [\mnras] {10.1093/mnras/stu2284}, \href
  {http://cdsads.u-strasbg.fr/abs/2015MNRAS.446.2380B} {446, 2380}

\bibitem[\protect\citeauthoryear{{Behrend} \& {Maeder}}{{Behrend} \&
  {Maeder}}{2001}]{behrend2001}
{Behrend} R.,  {Maeder} A.,  2001, \mn@doi [\aap] {10.1051/0004-6361:20010585},
  \href {http://cdsads.u-strasbg.fr/abs/2001A%26A...373..190B} {373, 190}

\bibitem[\protect\citeauthoryear{{Bernasconi} \& {Maeder}}{{Bernasconi} \&
  {Maeder}}{1996}]{bernasconi1996a}
{Bernasconi} P.~A.,  {Maeder} A.,  1996, \aap, \href
  {http://cdsads.u-strasbg.fr/abs/1996A%26A...307..829B} {307, 829}

\bibitem[\protect\citeauthoryear{{Chandrasekhar}}{{Chandrasekhar}}{1964}]{chan%
drasekhar1964}
{Chandrasekhar} S.,  1964, \mn@doi [\apj] {10.1086/147938}, \href
  {http://cdsads.u-strasbg.fr/abs/1964ApJ...140..417C} {140, 417}

\bibitem[\protect\citeauthoryear{{Dijkstra}, {Ferrara}  \&
  {Mesinger}}{{Dijkstra} et~al.}{2014}]{dijkstra2014}
{Dijkstra} M.,  {Ferrara} A.,   {Mesinger} A.,  2014, \mn@doi [\mnras]
  {10.1093/mnras/stu1007}, \href
  {http://cdsads.u-strasbg.fr/abs/2014MNRAS.442.2036D} {442, 2036}

\bibitem[\protect\citeauthoryear{{Eggenberger}, {Meynet}, {Maeder}, {Hirschi},
  {Charbonnel}, {Talon}  \& {Ekstr{\"o}m}}{{Eggenberger}
  et~al.}{2008}]{eggenberger2008}
{Eggenberger} P.,  {Meynet} G.,  {Maeder} A.,  {Hirschi} R.,  {Charbonnel} C.,
  {Talon} S.,   {Ekstr{\"o}m} S.,  2008, \mn@doi [\apss]
  {10.1007/s10509-007-9511-y}, \href
  {http://cdsads.u-strasbg.fr/abs/2008Ap%26SS.316...43E} {316, 43}

\bibitem[\protect\citeauthoryear{{Fricke}}{{Fricke}}{1973}]{fricke1973}
{Fricke} K.~J.,  1973, \mn@doi [\apj] {10.1086/152280}, \href
  {http://cdsads.u-strasbg.fr/abs/1973ApJ...183..941F} {183, 941}

\bibitem[\protect\citeauthoryear{{Fuller}, {Woosley}  \& {Weaver}}{{Fuller}
  et~al.}{1986}]{fuller1986}
{Fuller} G.~M.,  {Woosley} S.~E.,   {Weaver} T.~A.,  1986, \mn@doi [\apj]
  {10.1086/164452}, \href {http://cdsads.u-strasbg.fr/abs/1986ApJ...307..675F}
  {307, 675}

\bibitem[\protect\citeauthoryear{{Haemmerl{\'e}}, {Eggenberger}, {Meynet},
  {Maeder}  \& {Charbonnel}}{{Haemmerl{\'e}} et~al.}{2013}]{haemmerle2013}
{Haemmerl{\'e}} L.,  {Eggenberger} P.,  {Meynet} G.,  {Maeder} A.,
  {Charbonnel} C.,  2013, \mn@doi [\aap] {10.1051/0004-6361/201321359}, \href
  {http://cdsads.u-strasbg.fr/abs/2013A%26A...557A.112H} {557, A112}

\bibitem[\protect\citeauthoryear{{Haemmerl{\'e}}, {Eggenberger}, {Meynet},
  {Maeder}  \& {Charbonnel}}{{Haemmerl{\'e}} et~al.}{2016}]{haemmerle2016a}
{Haemmerl{\'e}} L.,  {Eggenberger} P.,  {Meynet} G.,  {Maeder} A.,
  {Charbonnel} C.,  2016, \mn@doi [\aap] {10.1051/0004-6361/201527202}, \href
  {http://cdsads.u-strasbg.fr/abs/2016A%26A...585A..65H} {585, A65}

\bibitem[\protect\citeauthoryear{{Hartwig} et~al.,}{{Hartwig}
  et~al.}{2016}]{hartwig2016}
{Hartwig} T.,  et~al., 2016, \mn@doi [\mnras] {10.1093/mnras/stw1775}, \href
  {http://cdsads.u-strasbg.fr/abs/2016MNRAS.462.2184H} {462, 2184}

\bibitem[\protect\citeauthoryear{{Hirano}, {Hosokawa}, {Yoshida}, {Umeda},
  {Omukai}, {Chiaki}  \& {Yorke}}{{Hirano} et~al.}{2014}]{hirano2014}
{Hirano} S.,  {Hosokawa} T.,  {Yoshida} N.,  {Umeda} H.,  {Omukai} K.,
  {Chiaki} G.,   {Yorke} H.~W.,  2014, \mn@doi [\apj]
  {10.1088/0004-637X/781/2/60}, \href
  {http://cdsads.u-strasbg.fr/abs/2014ApJ...781...60H} {781, 60}

\bibitem[\protect\citeauthoryear{{Hosokawa}, {Yorke}  \& {Omukai}}{{Hosokawa}
  et~al.}{2010}]{hosokawa2010}
{Hosokawa} T.,  {Yorke} H.~W.,   {Omukai} K.,  2010, \mn@doi [\apj]
  {10.1088/0004-637X/721/1/478}, \href
  {http://cdsads.u-strasbg.fr/abs/2010ApJ...721..478H} {721, 478}

\bibitem[\protect\citeauthoryear{{Hosokawa}, {Omukai}, {Yoshida}  \&
  {Yorke}}{{Hosokawa} et~al.}{2011}]{hosokawa2011b}
{Hosokawa} T.,  {Omukai} K.,  {Yoshida} N.,   {Yorke} H.~W.,  2011, \mn@doi
  [Science] {10.1126/science.1207433}, \href
  {http://cdsads.u-strasbg.fr/abs/2011Sci...334.1250H} {334, 1250}

\bibitem[\protect\citeauthoryear{{Hosokawa}, {Omukai}  \& {Yorke}}{{Hosokawa}
  et~al.}{2012}]{hosokawa2012a}
{Hosokawa} T.,  {Omukai} K.,   {Yorke} H.~W.,  2012, \mn@doi [\apj]
  {10.1088/0004-637X/756/1/93}, \href
  {http://cdsads.u-strasbg.fr/abs/2012ApJ...756...93H} {756, 93}

\bibitem[\protect\citeauthoryear{{Hosokawa}, {Yorke}, {Inayoshi}, {Omukai}  \&
  {Yoshida}}{{Hosokawa} et~al.}{2013}]{hosokawa2013}
{Hosokawa} T.,  {Yorke} H.~W.,  {Inayoshi} K.,  {Omukai} K.,   {Yoshida} N.,
  2013, \mn@doi [\apj] {10.1088/0004-637X/778/2/178}, \href
  {http://cdsads.u-strasbg.fr/abs/2013ApJ...778..178H} {778, 178}

\bibitem[\protect\citeauthoryear{{Iben}}{{Iben}}{1963}]{iben1963}
{Iben} Jr. I.,  1963, \mn@doi [\apj] {10.1086/147708}, \href
  {http://cdsads.u-strasbg.fr/abs/1963ApJ...138.1090I} {138, 1090}

\bibitem[\protect\citeauthoryear{{Iglesias} \& {Rogers}}{{Iglesias} \&
  {Rogers}}{1996}]{iglesias1996}
{Iglesias} C.~A.,  {Rogers} F.~J.,  1996, \mn@doi [\apj] {10.1086/177381},
  \href {http://cdsads.u-strasbg.fr/abs/1996ApJ...464..943I} {464, 943}

\bibitem[\protect\citeauthoryear{{Inayoshi} \& {Haiman}}{{Inayoshi} \&
  {Haiman}}{2014}]{inayoshi2014b}
{Inayoshi} K.,  {Haiman} Z.,  2014, \mn@doi [\mnras] {10.1093/mnras/stu1870},
  \href {http://cdsads.u-strasbg.fr/abs/2014MNRAS.445.1549I} {445, 1549}

\bibitem[\protect\citeauthoryear{{Inayoshi} \& {Omukai}}{{Inayoshi} \&
  {Omukai}}{2012}]{inayoshi2012}
{Inayoshi} K.,  {Omukai} K.,  2012, \mn@doi [\mnras]
  {10.1111/j.1365-2966.2012.20812.x}, \href
  {http://cdsads.u-strasbg.fr/abs/2012MNRAS.422.2539I} {422, 2539}

\bibitem[\protect\citeauthoryear{{Inayoshi}, {Visbal}  \&
  {Kashiyama}}{{Inayoshi} et~al.}{2015}]{inayoshi2015b}
{Inayoshi} K.,  {Visbal} E.,   {Kashiyama} K.,  2015, \mn@doi [\mnras]
  {10.1093/mnras/stv1654}, \href
  {http://cdsads.u-strasbg.fr/abs/2015MNRAS.453.1692I} {453, 1692}

\bibitem[\protect\citeauthoryear{{Inayoshi}, {Haiman}  \&
  {Ostriker}}{{Inayoshi} et~al.}{2016}]{inayoshi2016a}
{Inayoshi} K.,  {Haiman} Z.,   {Ostriker} J.~P.,  2016, \mn@doi [\mnras]
  {10.1093/mnras/stw836}, \href
  {http://cdsads.u-strasbg.fr/abs/2016MNRAS.459.3738I} {459, 3738}

\bibitem[\protect\citeauthoryear{{Johnson}, {Whalen}, {Fryer}  \&
  {Li}}{{Johnson} et~al.}{2012}]{johnson2012}
{Johnson} J.~L.,  {Whalen} D.~J.,  {Fryer} C.~L.,   {Li} H.,  2012, \mn@doi
  [\apj] {10.1088/0004-637X/750/1/66}, \href
  {http://cdsads.u-strasbg.fr/abs/2012ApJ...750...66J} {750, 66}

\bibitem[\protect\citeauthoryear{{Larson}}{{Larson}}{1972}]{larson1972}
{Larson} R.~B.,  1972, \mnras, \href
  {http://cdsads.u-strasbg.fr/abs/1972MNRAS.157..121L} {157, 121}

\bibitem[\protect\citeauthoryear{{Latif}, {Schleicher}, {Schmidt}  \&
  {Niemeyer}}{{Latif} et~al.}{2013a}]{latif2013d}
{Latif} M.~A.,  {Schleicher} D.~R.~G.,  {Schmidt} W.,   {Niemeyer} J.,  2013a,
  \mn@doi [\mnras] {10.1093/mnras/stt834}, \href
  {http://cdsads.u-strasbg.fr/abs/2013MNRAS.433.1607L} {433, 1607}

\bibitem[\protect\citeauthoryear{{Latif}, {Schleicher}, {Schmidt}  \&
  {Niemeyer}}{{Latif} et~al.}{2013b}]{latif2013e}
{Latif} M.~A.,  {Schleicher} D.~R.~G.,  {Schmidt} W.,   {Niemeyer} J.~C.,
  2013b, \mn@doi [\mnras] {10.1093/mnras/stt1786}, \href
  {http://cdsads.u-strasbg.fr/abs/2013MNRAS.436.2989L} {436, 2989}

\bibitem[\protect\citeauthoryear{{Mortlock} et~al.,}{{Mortlock}
  et~al.}{2011}]{mortlock2011}
{Mortlock} D.~J.,  et~al., 2011, \mn@doi [\nat] {10.1038/nature10159}, \href
  {http://cdsads.u-strasbg.fr/abs/2011Natur.474..616M} {474, 616}

\bibitem[\protect\citeauthoryear{{Omukai} \& {Palla}}{{Omukai} \&
  {Palla}}{2001}]{omukai2001}
{Omukai} K.,  {Palla} F.,  2001, \mn@doi [\apjl] {10.1086/324410}, \href
  {http://cdsads.u-strasbg.fr/abs/2001ApJ...561L..55O} {561, L55}

\bibitem[\protect\citeauthoryear{{Omukai} \& {Palla}}{{Omukai} \&
  {Palla}}{2003}]{omukai2003}
{Omukai} K.,  {Palla} F.,  2003, \mn@doi [\apj] {10.1086/374810}, \href
  {http://cdsads.u-strasbg.fr/abs/2003ApJ...589..677O} {589, 677}

\bibitem[\protect\citeauthoryear{{Osaki}}{{Osaki}}{1966}]{osaki1966}
{Osaki} Y.,  1966, \pasj, \href
  {http://cdsads.u-strasbg.fr/abs/1966PASJ...18..384O} {18, 384}

\bibitem[\protect\citeauthoryear{{Palla} \& {Stahler}}{{Palla} \&
  {Stahler}}{1992}]{palla1992}
{Palla} F.,  {Stahler} S.~W.,  1992, \mn@doi [\apj] {10.1086/171468}, \href
  {http://cdsads.u-strasbg.fr/abs/1992ApJ...392..667P} {392, 667}

\bibitem[\protect\citeauthoryear{{Pallottini} et~al.,}{{Pallottini}
  et~al.}{2015}]{pallottini2015}
{Pallottini} A.,  et~al., 2015, \mn@doi [\mnras] {10.1093/mnras/stv1795}, \href
  {http://cdsads.u-strasbg.fr/abs/2015MNRAS.453.2465P} {453, 2465}

\bibitem[\protect\citeauthoryear{{Park} \& {Ricotti}}{{Park} \&
  {Ricotti}}{2011}]{park2011}
{Park} K.,  {Ricotti} M.,  2011, \mn@doi [\apj] {10.1088/0004-637X/739/1/2},
  \href {http://cdsads.u-strasbg.fr/abs/2011ApJ...739....2P} {739, 2}

\bibitem[\protect\citeauthoryear{{Pezzulli}, {Valiante}  \&
  {Schneider}}{{Pezzulli} et~al.}{2016}]{pezzulli2016}
{Pezzulli} E.,  {Valiante} R.,   {Schneider} R.,  2016, \mn@doi [\mnras]
  {10.1093/mnras/stw505}, \href
  {http://cdsads.u-strasbg.fr/abs/2016MNRAS.458.3047P} {458, 3047}

\bibitem[\protect\citeauthoryear{{Regan}, {Johansson}  \& {Haehnelt}}{{Regan}
  et~al.}{2014}]{regan2014}
{Regan} J.~A.,  {Johansson} P.~H.,   {Haehnelt} M.~G.,  2014, \mn@doi [\mnras]
  {10.1093/mnras/stu068}, \href
  {http://cdsads.u-strasbg.fr/abs/2014MNRAS.439.1160R} {439, 1160}

\bibitem[\protect\citeauthoryear{{Sakurai}, {Hosokawa}, {Yoshida}  \&
  {Yorke}}{{Sakurai} et~al.}{2015}]{sakurai2015}
{Sakurai} Y.,  {Hosokawa} T.,  {Yoshida} N.,   {Yorke} H.~W.,  2015, \mn@doi
  [\mnras] {10.1093/mnras/stv1346}, \href
  {http://cdsads.u-strasbg.fr/abs/2015MNRAS.452..755S} {452, 755}

\bibitem[\protect\citeauthoryear{{Sakurai}, {Inayoshi}  \& {Haiman}}{{Sakurai}
  et~al.}{2016}]{sakurai2016b}
{Sakurai} Y.,  {Inayoshi} K.,   {Haiman} Z.,  2016, \mn@doi [\mnras]
  {10.1093/mnras/stw1652}, \href
  {http://cdsads.u-strasbg.fr/abs/2016MNRAS.461.4496S} {461, 4496}

\bibitem[\protect\citeauthoryear{{Schaerer}}{{Schaerer}}{2002}]{schaerer2002}
{Schaerer} D.,  2002, \mn@doi [\aap] {10.1051/0004-6361:20011619}, \href
  {http://cdsads.u-strasbg.fr/abs/2002A%26A...382...28S} {382, 28}

\bibitem[\protect\citeauthoryear{{Schleicher}, {Palla}, {Ferrara}, {Galli}  \&
  {Latif}}{{Schleicher} et~al.}{2013}]{schleicher2013}
{Schleicher} D.~R.~G.,  {Palla} F.,  {Ferrara} A.,  {Galli} D.,   {Latif} M.,
  2013, \mn@doi [\aap] {10.1051/0004-6361/201321949}, \href
  {http://cdsads.u-strasbg.fr/abs/2013A%26A...558A..59S} {558, A59}

\bibitem[\protect\citeauthoryear{{Shapiro} \& {Teukolsky}}{{Shapiro} \&
  {Teukolsky}}{1979}]{shapiro1979}
{Shapiro} S.~L.,  {Teukolsky} S.~A.,  1979, \mn@doi [\apjl] {10.1086/183134},
  \href {http://cdsads.u-strasbg.fr/abs/1979ApJ...234L.177S} {234, L177}

\bibitem[\protect\citeauthoryear{{Smidt}, {Whalen}, {Johnson}  \& {Li}}{{Smidt}
  et~al.}{2017}]{smidt2017}
{Smidt} J.,  {Whalen} D.~J.,  {Johnson} J.~L.,   {Li} H.,  2017,
  arXiv:1703.00449, \href {http://adsabs.harvard.edu/abs/2017arXiv170300449S}
  {}

\bibitem[\protect\citeauthoryear{{Sobral}, {Matthee}, {Darvish}, {Schaerer},
  {Mobasher}, {R{\"o}ttgering}, {Santos}  \& {Hemmati}}{{Sobral}
  et~al.}{2015}]{sobral2015}
{Sobral} D.,  {Matthee} J.,  {Darvish} B.,  {Schaerer} D.,  {Mobasher} B.,
  {R{\"o}ttgering} H.~J.~A.,  {Santos} S.,   {Hemmati} S.,  2015, \mn@doi
  [\apj] {10.1088/0004-637X/808/2/139}, \href
  {http://cdsads.u-strasbg.fr/abs/2015ApJ...808..139S} {808, 139}

\bibitem[\protect\citeauthoryear{{Umeda}, {Hosokawa}, {Omukai}  \&
  {Yoshida}}{{Umeda} et~al.}{2016}]{umeda2016}
{Umeda} H.,  {Hosokawa} T.,  {Omukai} K.,   {Yoshida} N.,  2016, \mn@doi
  [\apjl] {10.3847/2041-8205/830/2/L34}, \href
  {http://cdsads.u-strasbg.fr/abs/2016ApJ...830L..34U} {830, L34}

\bibitem[\protect\citeauthoryear{{Unno}}{{Unno}}{1971}]{unno1971}
{Unno} W.,  1971, \pasj, \href
  {http://cdsads.u-strasbg.fr/abs/1971PASJ...23..123U} {23, 123}

\bibitem[\protect\citeauthoryear{{Volonteri}, {Silk}  \& {Dubus}}{{Volonteri}
  et~al.}{2015}]{volonteri2015}
{Volonteri} M.,  {Silk} J.,   {Dubus} G.,  2015, \mn@doi [\apj]
  {10.1088/0004-637X/804/2/148}, \href
  {http://adsabs.harvard.edu/abs/2015ApJ...804..148V} {804, 148}

\bibitem[\protect\citeauthoryear{{Whalen} \& {Fryer}}{{Whalen} \&
  {Fryer}}{2012}]{whalen2012}
{Whalen} D.~J.,  {Fryer} C.~L.,  2012, \mn@doi [\apjl]
  {10.1088/2041-8205/756/1/L19}, \href
  {http://cdsads.u-strasbg.fr/abs/2012ApJ...756L..19W} {756, L19}

\bibitem[\protect\citeauthoryear{{Whalen}, {Abel}  \& {Norman}}{{Whalen}
  et~al.}{2004}]{whalen2004}
{Whalen} D.,  {Abel} T.,   {Norman} M.~L.,  2004, \mn@doi [\apj]
  {10.1086/421548}, \href {http://cdsads.u-strasbg.fr/abs/2004ApJ...610...14W}
  {610, 14}

\bibitem[\protect\citeauthoryear{{Woods}, {Heger}, {Whalen}, {Haemmerle}  \&
  {Klessen}}{{Woods} et~al.}{2017}]{woods2017}
{Woods} T.~E.,  {Heger} A.,  {Whalen} D.~J.,  {Haemmerle} L.,   {Klessen}
  R.~S.,  2017, preprint, \href
  {http://cdsads.u-strasbg.fr/abs/2017arXiv170307480W} {} (\mn@eprint {arXiv}
  {1703.07480})

\bibitem[\protect\citeauthoryear{{Wu} et~al.,}{{Wu} et~al.}{2015}]{wu2015}
{Wu} X.-B.,  et~al., 2015, \mn@doi [\nat] {10.1038/nature14241}, \href
  {http://cdsads.u-strasbg.fr/abs/2015Natur.518..512W} {518, 512}

\bibitem[\protect\citeauthoryear{{Yorke} \& {Bodenheimer}}{{Yorke} \&
  {Bodenheimer}}{2008}]{yorke2008}
{Yorke} H.~W.,  {Bodenheimer} P.,  2008, in {Beuther} H.,  {Linz} H.,
  {Henning} T.,  eds,  Astronomical Society of the Pacific Conference Series
  Vol. 387, Massive Star Formation: Observations Confront Theory. p.~189

\makeatother
\end{thebibliography}

\bsp	
\label{lastpage}
\end{document}